\documentclass[usenatbib,useAMS]{mn2e}
\usepackage{graphicx}
\usepackage{times}
\usepackage{multirow}
\usepackage{bm}
\usepackage{bib_macro}
\newcommand{\be}{\begin{equation}}
\newcommand{\ee}{\end{equation}}
\newcommand{\ba}{\begin{eqnarray}}
\newcommand{\ea}{\end{eqnarray}}

\newcommand{\rr}{{RRLyr\ae}}
\newcommand{\rrs}{{RRLyr\ae s}}
\newcommand{\ft}{\bm {{\cal F}}}
\newcommand{\ftcc}{\bm {{\bar  {\cal F}}}}
\newcommand{\ift}{\bm {{\cal  F}^{-1}}}

\newcommand{\mathsmall}[1]{\mbox{\small$#1$}}

\def\gs{\mathrel{\raise1.16pt\hbox{$>$}\kern-7.0pt %  >
\lower3.06pt\hbox{{$\scriptstyle \sim$}}}}         %  ~
\def\ls{\mathrel{\raise1.16pt\hbox{$<$}\kern-7.0pt %  <
\lower3.06pt\hbox{{$\scriptstyle \sim$}}}}         %  ~
\def\registered{{\ooalign{\hfil\raise .00ex\hbox{\scriptsize R}\hfil\crcr\mathhexbox20D}}}

\suppressfloats

\begin{document}
\bibliographystyle{mn2e}

\title[Finding  outlier light-curves in catalogs of periodic variable stars]{Finding outlier light-curves in catalogs of periodic variable stars}
\author[Protopapas et al. ]
{P.~Protopapas$^1$, J.~M.~Giammarco$^2$, L.~Faccioli$^2$, M.~F.~Struble$^2$,R.~Dave$^2$, C.~Alcock$^1$\\
$^1$Harvard Smithsonian Center for Astrophysics, 60 Garden Street, Cambridge, MA 02138, USA.\\
$^2$Dept. of Physics and Astronomy, University of Pennsylvania, Philadelphia,
PA 19104, USA.}
\maketitle

\begin{abstract}
We present a methodology to discover outliers in  catalogs of
periodic light-curves. We use cross-correlation as measure of
``similarity'' between two individual light-curves and then
classify light-curves with lowest average ``similarity'' as
outliers. We performed the analysis on catalogs of variable stars
of known type from the MACHO and OGLE projects and established
that our method correctly identifies light-curves that do not
belong to those catalogs as outliers. We show how our method can
scale to large datasets that will be available in the near future
such as those anticipated from Pan-STARRS and LSST.

\end{abstract}


% =============================================================
% =============================================================
%    INTRODUCTION
% =============================================================
% =============================================================
\section{Introduction}
One major byproduct of the completed MACHO and ongoing OGLE, EROS,
and MOA microlensing surveys are catalogs of $\sim10^5$ variable
stars generated from long temporal photometric monitoring of stars
in selected fields of the Magellanic Clouds and the Galactic bulge
\citep{Ferlet1997, Paczynski2001}. Most of these are comprised of
periodic variable stars, whose periods were estimated via various
statistical techniques \citep{Lomb1976, reimann94}, and a
smaller number are comprised of non-periodic variable stars.  Periodic
variable stars have been classified by eye, based primarily on the
visual appearance of their light-curves folded with an estimated
period,
 and their locations in the color-magnitude and period-luminosity diagrams.
Automatic procedures are available using Fourier coefficients
\citep{Morgan1998} and neural networks \citep{Belokurov2003,
Eyer2005}, and others are under development \citep{Wozniak2002}.
The reliability of type classification of light-curves with these
automated techniques is estimated to be $\sim90$\% \citep{Wozniak2002}.

A natural question that arises concerns the detection of outliers
in variable star catalogs, i.e., members whose light-curves
deviate at a prescribed statistical level from the rest.  There could
be several reasons for this: a poor or incorrect period caused by
 noisy photometric data, outright misclassification, or, perhaps
rarely and more interestingly, an intrinsic physical difference
such as a slowly changing period or brightness amplitude which
introduces noise in  the folded light-curve, analogous to the
longer term variability of the Cepheid variable Polaris
\citep{Evans2002,Engle2004}, or apsidal motion in eccentric
eclipsing binaries \citep{Wolf2001, Wolf2004}. While catalog
membership is nearly complete for variable stars derived from the
MACHO and OGLE projects, the growth of massive databases of
variable stars at fainter magnitudes is anticipated
\citep{Paczynski2001}, largely using automated procedures in
tandem with data-mining \citep{Belokurov2003}. This circumstance
recommends the development of a fast, reliable procedure to eliminate
contaminating outliers, so they may be subject to later review,
analysis, and reclassification.  Developing such a procedure to
find outliers in large datasets of variable stars provides the
motivation for the methodology described in this paper.

This paper is organized in the following way. Section~2 is devoted
to the methodology. In Section~3 we show how our method can be
extended to a large number of light-curves. In Section~4 we
present the results from runs on MACHO and OGLE catalogs.  Future
work is presented in Section~5 and conclusions are in Section~6.

%=====================================================================
%=====================================================================
%        METHODOLOGY
%=====================================================================
%=====================================================================
\section{Methodology}\label{sec:method} 
Our main objective is to identify outliers in a dataset of variable 
stars.  
The basic procedure is conceptually straightforward; compare the light-curve(s) 
 in the
dataset with that of every other light-curve in the dataset, and
see which light-curve(s) is least like all others.  Closer
scrutiny reveals some of the difficulties of this process. First,
given the size of the datasets ($\sim10^5$ for existing datasets,
growing to $\sim 10^8$ in the near future) the comparison
method(s) must be fast and scale favorably. Second, the size of
the datasets also prohibits human supervision, so the methods must
be automated and very robust.

\begin{figure}
\begin{center}
\includegraphics[width=2.0in]{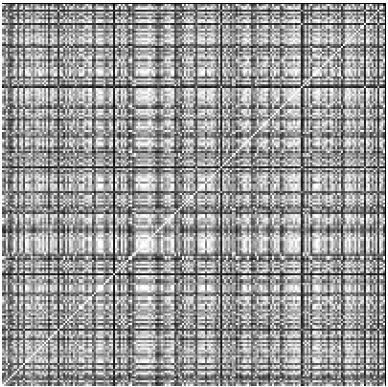}
\caption{\label{Fig:SimMatrix} {\small Similarity matrix for 200
light-curves. Each point represents the square of the correlation
between two light-curves. Bright points correspond to strongly
correlated/anti-correlated  pairs of light-curves. Dark points
correspond to weakly correlated pairs of light-curves. Note that
the indexing starts at the lower left hand corner with (1,1).}}
\end{center}
\end{figure}

Finding an outlier requires two separate comparisons. The first
comparison is between two individual light-curves to determine how
similar, or dissimilar, they are to each other.  This comparison
will be described in Section~\ref{Sec:RefSecLCComp}.  Once this
comparison is done for every pair of light-curves in the dataset
we form a similarity matrix (see Fig.~\ref{Fig:SimMatrix}). Each
row of the similarity matrix represents the similarity of a given
light-curve to all other light-curves in the dataset.  To
determine which light-curve in the dataset is least like all
others we compare the rows of the similarity matrix and determine
which row has on average the smallest similarity with every other
light-curve.  This second comparison is described in more detail
in Section~\ref{Sec:RefRowComp}.

We begin by describing  the preprocessing of the light-curves, and then the
actual comparison tests.

%

% ----------------------------------------------------------------
%     PREPROCESSING
% ----------------------------------------------------------------
\subsection{Preprocessing}

There is no one-shot approach to preprocessing a dataset of
light-curves. A smoothing technique will remove undesired noise
but could also remove true features of the light-curve. An
interpolation may generate a more natural looking light-curve but
can also insert features that are not physical. Sophisticated
signal processing methods can be used to determine the best
smoothing/interpolation/designaling method, however this will only
be true for a single light-curve. Since we are dealing with a
large collection of light-curves and essentially we are looking
for a few different light-curves, using a universal
preprocessing algorithm is not a sensible strategy. For these
reasons, we have chosen a minimal preprocessing scheme; one  that
preserves the main light-curve features but does not allow obvious
spikes to dominate the statistics.\footnote{Here statistics refers
to the overall outlier measure which is described in Section~\ref{Sec:RefRowComp}}

The steps that are described below in this section are the steps
used for the analysis done in Section~\ref{sec:results} on the
MACHO and OGLE catalogs. We have however experimented with a
number of different schemes and the resulting modules developed
will be released as part of the software suite. We have concluded
that while the comparisons between pairs of light-curves do depend
on the choice of preprocessing scheme the measure of overall
outlier does not closely depend on the choice of parameters used
in the preprocessing or the preprocessing method (assuming we stay
within reasonable limits).

 \label{Sec:Preprocessing} For any measure of
similarity to be meaningful, the light-curves must be preprocessed
to  retain  the true features of the data, while minimizing the
effects of noise and spurious measurements.
 Currently our comparison methods require the
values of the light-curves at predetermined, uniformly spaced,
times.\footnote{Our current FFT method requires measurements
uniformly spaced in time.  Additionally, any time domain
comparison method would require knowing the measurements at
predetermined times.} Since we need the values of the light-curves
at uniformly spaced intervals we need to interpolate the 
light-curves.  All light-curves have spurious data due to noise
and other effects, and many have spikes.

Any interpolation method may be adversely affected by these spikes
and high-frequency noise. For this reason we have built into our
methodology a three step
spike-removal/interpolation/data-smoothing process. We first
perform a running average on the light-curve data (spike-removal),
we then perform an interpolation to obtain the values of the
light-curve at prescribed times.  We then perform a smoothing
process on the interpolated data.  This smoothing process is a
generalized Savitzky-Golay (SG) smoothing \citep{Gorry1990}.

\underline{Running average:} Our running average scheme replaces
the value of each data point by the average of the data points
contained within a box centered on the data point.  Since our data
are not evenly spaced we weigh the influence any value can have on
the running average by its distance to the box center.  We use a
Gaussian weight that depends on distance from the ``current
point'' and has a standard deviation  half the window size.  The
results of a running average are somewhat dependent on the width
of the running window size. Since we wish to remove spikes but
not features, we determined that a width of 1\% of the light-curve
phase worked well. An extension to this method is to additionally
weigh the values by the observational error using Gaussian
weights. This modification turned out to be extremely useful in
very large datasets where observational errors cannot be
accounted in the measure of correlation. This point will become
clearer in the following sections.

\underline{Interpolation:} We use simple linear
interpolation in order to produce uniformly spaced light-curve
points.  We have found that a linear interpolation, in combination
with the spike-removal and the smoothing, described next, works
well in practice.

\underline{Smoothing:} The post-interpolation smoothing method
uses a generalized Savitzky-Golay method.  Savitzky-Golay is a
well known and widely used smoothing method \citep{NR}. The method
we employ is generalized because it does not truncate the
endpoints of the dataset in the smoothing process.  It does this
by employing the Gram polynomials.  A typical implementation of
the SG smoothing algorithm is, in a sense, a running least squares
fit to the data and requires solution of a matrix equation as we
march through the data.  Using the recursive properties of the
Gram polynomials, as in \cite{Gorry1990}, SG smoothing can
be accomplished without the need to solve matrix equations.

There are two adjustable parameters in our SG smoothing.  The
order of the polynomials, and the width of the smoothing window.
Since smoothing of the data is the principle objective of this
procedure, we typically use third order polynomials, attempting to
smooth out the higher order oscillations.  The width of the
smoothing window determines the range of influence a given point
has over neighboring points (the larger the window, the more
neighboring points affect the smoothed value of the current
point).  Not wanting to ``smooth-out'' any features we determined
that a width of 4\% of the period worked well.  A review
of the properties of Savitzky-Golay filters can be found in \cite{Luo2004}.

\begin{figure}
\begin{center}
\includegraphics[width=6.0cm]{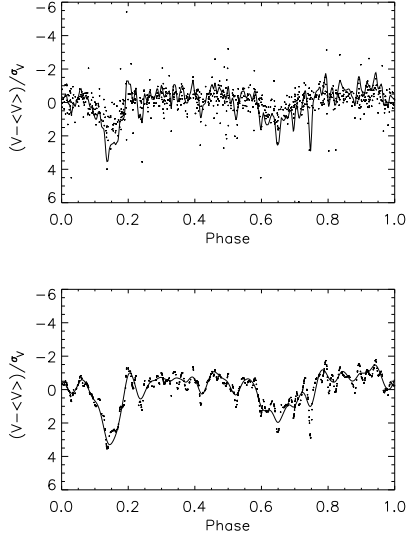}
\end{center}
\caption{A light-curve as it is passed through the pre-processing
steps. The points in the top panel shows the original
light-curve. The solid line in the same panel shows the
light-curve after the spike-removal is performed. The solid line
in the second panel show the final result after interpolation and
smoothing. In the same panel the results after spike-removal are
shown for comparison.  \label{Fig:PreProcSteps1} }
\end{figure}

Fig.~\ref{Fig:PreProcSteps1} shows the modifications in a given
to a folded light-curve as it is passed through the pre-processing steps
described above. The points in the top panel shows the original
light-curve. The solid line in the same panel shows the
light-curve after the spike-removal is performed. The solid line
in the second panel shows the final result after interpolation and
smoothing. In the same panel the results after spike-removal are
shown for comparison. Upon inspection of Fig.~\ref{Fig:PreProcSteps1}
one will notice that the differences between the initial,
pre spike-removal light-curve and final smoothed light-curve
is perhaps not as dramatic as could be
achieved, or that more smoothing could have been accomplished in
the spike-removal stage. While this is true we preferred to err
on the side of caution, resisting the temptation to produce very
smooth light-curves while being certain to preserve features
within the light-curve.

Note that at each preprocessing step we have estimated the errors
using typical error propagations techniques (see Appendix
\ref{app:errors} for details). Hence the final light-curve
contains observational errors that are  necessary for the
next stage.

%============================================================================
%
%          Figure preprocessing
%
%============================================================================

%-------------------------------------------------------------------------
% Comparison
%-------------------------------------------------------------------------

\subsection{Comparison of Light-Curves} \label{Sec:RefSecLCComp}

For most tests, a comparison of two light-curves is a
point-by-point comparison of two time series.  In this work we
have concentrated on the use of the correlation between two
light-curves as the measure of their ``similarity''. There are
many choices of measure of similarity and depending on the
``features'' of the light-curves some work better than others.
Cross-correlation and chi-square tests are the simplest choices.
One can show though, that the order of outliers remains the same
nonetheless.  Future work will investigate different measures of
similarity.

%============================================================================
% Correlation coefficient
%============================================================================
\subsubsection{Correlation Coefficient of two time series with
measurement errors} \label{Sec:RefSecCorrCoef} The uncertainties
in the flux measurements of a typical light-curve can vary
significantly.  For this reason any analysis based on the flux
must account for the variations in the uncertainties in the flux
measurements.

The goal is to derive a modified correlation coefficient $r$ of
two light-curves that incorporates the errors of the measurements.

We begin by considering  the  ``standard'' correlation coefficient (without
observational errors) of two times series $y(n)$ and $x(n)$ where
$n$ is the discrete time. For each measurement $y(n),x(n)$ there
are associated measurement errors $\sigma_{y}(n)$ and
$\sigma_{x}(n)$. For the moment we assume the averages of $y(n)$
and $x(n)$ to be zero.

We examine how well the data fit the line $y=\alpha x$. Using a
least square fit
\begin{equation}
    \chi^2 = \sum_n \left[ y(n) - \alpha x(n) \right] ^2 \: ,
\end{equation}
then by taking the derivative with respect to $\alpha$ we can show
that the $\chi^2$ is a minimum when

\begin{equation}
    \alpha = \frac{ \sum_n y(n) \, x(n) }
        { \sum_n x^2(n)  } .
\end{equation}
Performing a least squares fit on the inverse equation $x=\beta
y$, we can similarly show that
\begin{equation}
    \beta = \frac{ \sum_n y(n) \, x(n) }
        { \sum_n  y^2(n)  } \: .
\end{equation}
The correlation coefficient is defined as \citep{Weisstein1}:
\begin{equation}
    r_{xy} \equiv \sqrt {\alpha \beta} =   \frac{ \sum_n  y(n) x(n)  }
        {\sqrt{ \sum_n y^2(n)  \sum_n x^2(n)  } }
        \: .
\end{equation}

\noindent This is the correlation coefficient without
observational errors. In the case of observational errors, fitting
the linear equations $y=\alpha x$, $x=\beta y$ using a $\chi^2$
yields,
\begin{equation}
    \chi^2 = \sum_n \frac{ \left[ y(n) - \alpha x(n)\right] ^2 }{ \sigma^2_{y}(n)} \: .
\end{equation}
Setting the derivative with respect to $\alpha$ equal to zero we
can show that

\begin{equation}
    \alpha = \frac{
           \sum_n  y(n) x(n)  /
                            \sigma_{y}^{2}(n)
        }{  \sum_n x^{2}(n) / \sigma^{2}_y(n)} \: ,
\end{equation}
and equivalently
\begin{equation}
    \beta = \frac{
           \sum_n y(n) x(n)  /
                            \sigma_{x}^{2}(n)
        }{  \sum_n y^{2}(n) / \sigma^{2}_x(n)} \: .
\end{equation}
Using the above definition of the correlation coefficient we can
show
\begin{eqnarray}
    r_{xy} &=& \sqrt {\alpha \beta} = \nonumber \\
    & & \sqrt{\frac{ \sum_n y(n) x(n) / \sigma_y^{2}(n)
        \sum_n y(n) x(n)  / \sigma_x^{2}(n)
    }
        { \sum_{n}  x^2 (n)/  \sigma_y^{2}(n)  \sum_{n}  y^2(n) / \sigma_x^{2}(n)  }
    } \;.
\label{eq:corr_errors1}
\end{eqnarray}
If the mean values of $x$ and $y$ are not zero we can extend the
above analysis by using the following transformations,

\begin{eqnarray*}
  x'_{i} &\rightarrow& x_{i} - \bar{x}  \\
  y'_{i} &\rightarrow& y_{i} - \bar{y} .
\end{eqnarray*}
Substituting for the new variables in Eq.~\ref{eq:corr_errors1} we can show that
\begin{eqnarray}
\lefteqn{ \mathsmall{  r^{2}_{xy} = }} \nonumber \\
     && \!\!\!\!\!\!\!\!
   \mathsmall{  \frac{ \sum_{i} \left[ ( y(n)-\bar{y})( x(n)-\bar{x})  / \sigma_y^{2}(n) \right]
        \sum_n \left[ (y(n)-\bar{y})( x(n) -\bar{x}) / \sigma^2_{x}(n) \right]
    }
        { \sum_{n} \left[ (x(n)-\bar{x})^2 /  \sigma_y^{2}(n)\right] \sum_{n}
        \left[( y(n)-\bar{y})^2 / \sigma_{x}^{2}(n)  \right] }
        }
\label{eq:fullcorr}
\end{eqnarray}

%============================================================================
% Cross correlation in Fourier space
%============================================================================
\subsubsection{Cross correlation in Fourier space}
\label{Sec:Phasing} The comparison of two light-curves using the
correlation coefficient described above hinges on the choice of
epoch. Since the phase of the first signal  can be
arbitrarily chosen a comparison could yield a small $r^2$ even
if two light-curves are alike. Therefore this arbitrary
epoch has to be adjusted for all light-curves prior to any
comparison.

An obvious approach is to move the epoch of one of the two
light-curves until a maximum $r^2$ is achieved. Though
conceptually simple, this approach could be quite computationally
costly as it would need to be calculated for every pair of
light-curves. Fortunately, this can be performed quite
economically in Fourier space using the convolution theorem.

The correlation  between light-curve $x$ and light-curve $y$ with
time  lag  $\tau$ is given by
\begin{equation}
  r^2_{xy}(\tau) = \sum_{n=0}^{N-1} x(n) \, y(n-\tau) \: ,
\end{equation}
where $n$ is the discrete time.  According to the convolution
theorem (see Appendix \ref{app:conv}) the correlation can be
written as
\begin{equation}
\label{Equ:CorrelationFreqDomain}
  r^2_{xy}(\tau)
          =    {\cal F}^{-1}  \left[    {\cal
X}(\nu)  \, \bar{{\cal Y}}(\nu)    \right]  (\tau)
\end{equation}
where  ${\cal X}(\nu)$ is the Fourier transform of $x(n)$ and
$\bar{{\cal Y}}(\nu)$ is the complex conjugate of the Fourier
transform of $y(n)$. Therefore one can find the maximum
correlation by finding the maximum of the inverse Fourier
transform of the product of the Fourier transforms of the two
light-curves.  For fast Fourier transforms (FFT), each Fourier
transform  requires $2N \log(N)$ operations, where $N$ is the
number of observations. Thus for each pair of light-curves a total
of $6 N\log(N)$ operations are required. This is to be compared to
$N^2$ operations required doing the analysis in regular space.

The above equations can be extended to include measurement errors
(Eq.~\ref{eq:corr_errors1}).  \footnote{Finding the correlation of
Eq.~\ref{eq:fullcorr} will only require shifting the zeroth
component of the Fourier transforms.}

\begin{eqnarray}
    \label{eq:CorrelationFreqDomain}
 \lefteqn{
r^2_{xy}(\tau)   = } \\
& & \!\!\!\!\!\!\!\!\!\!\!\!\!\!  \frac{\ift \left[ \ft(
\frac{y(n)}{\sigma^2_y(n)}) \ftcc(x(n)) \right](\tau) \,\,\,
      \ift \left[ \ft(y(n))\ftcc( \frac{ x(n)}{\sigma^2_x(n)}) \right] (\tau)}
      { \ift \left[ \ft( \frac{1}{\sigma^2_y(n)}) \ftcc(x^2(n))\right] (\tau)
      \,\,\,  \ift \left[ \ft(\frac{1}{\sigma^2_x(n)}) \ftcc(y^2(n))\right]
      (\tau)} \nonumber \; .
\end{eqnarray}

\begin{figure}
\begin{center}
\includegraphics[width=6.5cm]{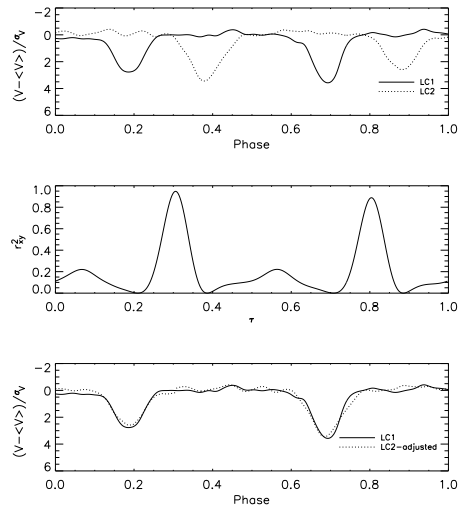}
\end{center}
\caption{Top panel shows two similar light-curves with arbitrary
epochs after being normalized and shifted to set the mode to
be one. The middle panel shows the square of the correlation plotted as
a function of the epoch. The maximum occurs at $\sim 0.3$. Finally
the bottom panel shows the two light-curves after one of the two
light-curves are time-shifted by $\sim0.3$.
\label{fig:convexample} }
\end{figure}

\noindent The top panel in Fig.~\ref{fig:convexample} shows two
light-curves with arbitrary epochs. The middle panel shows the
square of the correlation  as a function of the time lag,
$r^2_{xy}(\tau)$. The maximum occurs at $\tau \approx$ 0.3,
calculated using Eq.~\ref{eq:CorrelationFreqDomain}. The
bottom panel shows the same light-curves after one of the
light-curves is time-shifted by 0.3.

%============================================================================
%
%          Figure convolution example
%
%============================================================================

%\subsection{Correlation Vs. FFT Coefficients as a Comparison of Light-Curves}
%We are testing the possibility of using the frequency spectrum resulting from
%the FFT of the light-curves as a basis of comparison as opposed to their
%correlation.  That is, we compute the FFT of each light-curve, and compare the
%resulting frequency spectrums, perhaps using a Kolmogorov-Smirnov test.

%============================================================================
%       OUTLIER MEASURE
%============================================================================
\subsection{Outlier Measure} \label{Sec:RefRowComp} Once we have
completed the comparisons of each pair of light-curves, thus populating
the similarity matrix, we compare the rows of the similarity matrix
to determine the outliers. For each line in the similarity matrix
we compute  the average of the correlations as
\begin{equation}
  {\cal R}^2_x=  \frac{1}{N_{\mathrm{LC}}-1} \sum_{y \neq x} r^2_{xy}   \: ,
\end{equation}
where $y$ runs over all light-curves in the set except for $x$ and
$N_{\mathrm{LC}}$ is the number of light-curves.

For each light-curve we calculate the average of the correlations
as above and then we rank this measure. The light-curves with the
lowest correlation are classified as probable outliers and are
further inspected.
\par
How many  light-curves should be inspected? A natural choice is to
set a threshold  based on the actual value of the  average
correlation (${\cal R}$). For example we could set  the threshold
at ${\cal R}=0.3$; thus any light-curve below that value  should
be examined. Yet this is not exactly what we looking for. Consider
the following scenario:  a catalog consists of light-curves which
are all alike (e.g. a collection of well separated eclipsing
binary stars with circular orbits and components that are both O
stars). The light-curves of this collection will be naturally
strongly correlated. If one of the objects in the catalog is a
binary system with one of the stars being a B star then the
correlation to the rest of the light-curves will be slightly lower
but not low in absolute terms. Nevertheless that light-curve will
have the lowest ${\cal R}$ in the set thus should be flagged for
additional inspection. The same may apply to a collection of
light-curves that are classified together but their light-curves
show weak correlation (this is an indication that the band in
which the observations were made is not the primary manifestation
of the physical classification) therefore  a low correlation does
not necessarily mean that the particular light-curve is an
outlier. Hence, what really matters is the average correlation,
${\cal R}$, compared to the rest of ${\cal R}$'s in the set.

One could calculate the expectation value and variance of  the
distribution of ${\cal R}$'s and determined which light-curves are
at least 2$\sigma$'s away from the mean. This would have been  a
reasonable approach assuming the underling distribution was
Gaussian. Unfortunately this is not true in general. First
consider the case that all pairs of light-curves have the same correlation,
$\lambda$. The probability density function (pdf) of the
correlations of this set would be a bivariate normal distribution
(which at large $N$ becomes a Gaussian). 
%
%The mean and the moments
%of this distribution depend on the $\lambda$ as
%\begin{eqnarray}
%  \langle R \rangle &=& \lambda - \frac{\lambda(1-\lambda^2)}{2N} \\
%  \rm{var}(R) &=&  \frac{ (1-\lambda^2)^2}{N}\left( 1+ \frac{11 \lambda^2}{2N} + \ldots \right) \\
%  \gamma_1 &=& \frac{6 \lambda}{\sqrt{N}} \left( 1+ \frac{77 \lambda^2 -30}{12 n}   + \ldots \right)\\
%  \gamma_2 &=& \frac{6 }{N} \left( 12 \lambda^2 -1 \right) + \ldots
%\end{eqnarray}
%where $N$ is the number of observations. 
In reality our sets of light-curves do not all have the same correlations. For simplicity
assume that the light-curves could be grouped in clusters with
constant correlations. Then the resulting pdf will be a
superposition of bivariate normal distributions each with
different $\lambda$. Therefore the final pdf is  dataset dependent
and may or may not resemble a Gaussian. For these reasons, the
average of ${\cal R}$'s and its variance has proven not to be a
reliable approach. \footnote{We have tested the above empirically
and we found that in most cases the resulting pdf's are not
invariant.} Consequently we have concluded  that simply selecting
the light-curves with lowest average correlation (order of 5\%) is
the fastest and the most reliable approach.

%

%============================================================================
%============================================================================
%           LARGE DATASETS
%============================================================================
%============================================================================
\section{Large datasets}

The numerical method  shown in Section~\ref{sec:method} works well to
identify outliers. This will be  demonstrated in
Section~\ref{sec:results} where outliers are identified in real
light-curve catalogs.

%[NOT] We have also demonstrated that with the modular design of
%our package more comparisons can be tested and more outliers will
%be identified. Adding other comparisons and combining different
%comparison methods is the subject of new project underway.

For a  dataset containing $\sim 5000$ light-curves the run time on a
typical desktop (3~GHz Intel$^\registered$  Xeon\texttrademark)
is  $\sim 5$ hours.

\begin{figure}
\begin{center}
\includegraphics[width=7cm]{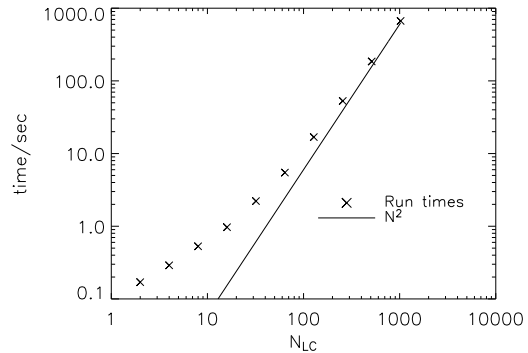}
\end{center}
\caption{Time in seconds to complete the analysis as a function of
number of light-curves on log-log scale. The points correspond to
the actual computational times while the solid line corresponds to
the $N^2$ relation. It is clear that for large $N$'s the
computational time scales as $N^2$. \label{Fig:Performace1} }
\end{figure}

The real advantage of a method like this would lie in the
ability to perform the analysis on much larger data-sets.
Unfortunately our method scales as  $N_{\mathrm{LC}}^2$ where
$N_{\mathrm{LC}}$ is the number of light-curves.
Fig.~\ref{Fig:Performace1} gives a graphical representation of the
performance of our model.  It shows running times, in seconds, as
a function of $N_{\mathrm{LC}}$ in a log-log scale. Superimposed
on this plot is a curve that is proportional to
$N_{\mathrm{LC}}^2$.  For large $N_{\mathrm{LC}}$  we see our
algorithm scales as $N^2_{\mathrm{LC}}$. Accordingly, for a
dataset of $\sim 10^5$ light-curves the analysis will take about
50 days! \footnote{ The program could be executed in parallel thus reducing
the computational time by a factor  of $n_\mathrm{cpu}$
($n_\mathrm{cpu}$ being the  number of cpu's). However the
datasets will soon grow to $10^6$ thus requiring few thousands of
cpu's in order to run the analysis in few days.} 
Consequently we must craft alternative, smarter algorithms
to deal with larger datasets.

In the following subsections we present alternative approaches to
speed up the calculation, each one having advantages and
disadvantages.
In section \ref{sec:simple}  we show how, in the
case of a simple correlation coefficient (without the observation
errors), the analysis in discovering outliers can  be reduced from
$N^2_\mathrm{LC}$ operations to  $N_\mathrm{LC}$ operations. In
Section~\ref{sec:subsets} we will show a simple approximation that
can be applied  to the extended correlation coefficient in
Eq.~\ref{eq:CorrelationFreqDomain} (including observational errors
and allowing time lag to vary).

%---------------------------------------------------
%  simple correlation for large datasets
%---------------------------------------------------

\subsection{Simple correlation coefficient}
\label{sec:simple} The correlation coefficient between two light-curves $i$,$j$ is given by
\begin{equation}
 r_{ij} = \frac{ \sum_{t} \left( f_i(t) - \bar{ f_i}\right) \left(f_j(t)-\bar{f_j}\right)}{(N-1) \, \sigma_i \sigma_j}
\end{equation}
where $N$ is the number of observations.
To identify outliers we calculated the average correlation of each
light-curve with the rest of the set (see
Sec.~\ref{Sec:RefRowComp}). This average correlation is given by
\begin{eqnarray}
  R_i &=& \frac{1}{N_{\mathrm{LC}}-1} \left(\sum_j r_{ij} - 1 \right) \\
   &=& \frac{1}{N_{\mathrm{LC}}-1} \left( \sum_j \left[ \frac{ \sum_{t} \left( f_i(t) - \bar{ f_i}\right) \left(f_j(t)-\bar{f_j}\right)}{(N-1) \, \sigma_i \, \sigma_j}\right] - 1 \right),\nonumber
\end{eqnarray}
where we sum over all $j$'s and subtract 1 for the $i=j$ case.
\par
\vspace{.3cm}
\noindent Re-arranging the order of the sums we get \footnote{Here we are making the assumption
that all light-curves have same $t$'s. This is not true in general but it is true
after proper interpolation -something we performed in the preprocessing steps. }
\begin{equation}
  R_i = \frac{1}{N_{\mathrm{LC}}-1} \left( \sum_t \left[ \frac{  \left( f_i(t) - \bar{ f_i}\right)}{(N-1) \, \sigma_i}  \sum_j \frac{\left(f_j(t)-\bar{f_j}\right)}{ \sigma_j}\right] - 1 \right)
\label{eq:cor1}
\end{equation}
We  define a centroid light-curve as:
\begin{equation}
    \label{eq:centroid}
  F(t) \equiv \frac{1}{N_{\mathrm{LC}}}\sum_j \frac{f_j(t)}{
  \sigma_j}\,\, ,
\end{equation}
and its average centroid light-curve
\begin{equation}
  \bar{F} = \frac{1}{N} \sum_t F(t)= \frac{1}{N_{\mathrm{LC}}}\sum_j \frac{\bar{f_j}}{\sigma_j}.
\end{equation}
Substituting the definitions of $F$ and $\bar{F}$ into Eq.~\ref{eq:cor1} we get
\begin{equation}
  R_i =  \frac{N_{\mathrm{LC}}}{N_{\mathrm{LC}}-1} \frac{\sum_{t}  \left( f_i(t) - \bar{ f_i}\right) \left(F(t)-\bar{F}\right)}{ (N-1) \, \sigma_i} -\frac{1}{N_{\mathrm{LC}}-1}.
\end{equation}
Note that at the limit where $N_{\mathrm{LC}}\gg 1$,
$\frac{N_{\mathrm{LC}}}{N_{\mathrm{LC}}-1} \rightarrow 1$ , $\frac{1}{N_{\mathrm{LC}}-1}
\rightarrow 0$ and therefore
\begin{equation}
  R_i =    \frac{ \sum_{t} \left( f_i(t) - \bar{ f_i}\right) \left(F(t)-\bar{F}\right)}{(N-1) \,
  \sigma_i}\,\,.
\end{equation}
Since $F(t)$ and $\bar{F}$ need to be calculated only once, the
number of operations necessary to find all $R_i$'s is
$O(N_{\mathrm{LC}}+N\times N_{\mathrm{LC}}) \sim O(N \times N_{\mathrm{LC}})$ which is a
significant improvement over the $O(N \times N_{\mathrm{LC}}^2)$ which was
necessary before.

This gain does not come without disadvantages. Firstly note that we
can {\bf not} apply the same transformation from
``average-of-the-correlations'' to ``correlation-to-the-average''
in the case of correlation coefficients using observational
errors, since in this case the magnitudes and the errors are
mixed. Nevertheless this is not a major disadvantage since the
observational errors can be partially taken into account in the
averaging/smoothing operations. The second major shortcoming is the fact
that the time lag cannot be considered as a free parameter. This
is because the time lag depends on both light-curves thus $F(t)$
cannot be calculated once for all light-curves. To circumvent
this problem we need to find  \emph{a priori} an \emph{absolute phase for all
light-curves}.

\subsection{Universal phasing}
To do just that  we have devised the following algorithm of
adjusting the epoch of all light-curves  using clustering methods.
The basic concept is to find where the signal with the
highest/lowest magnitude dip occurs  for each light-curve and set
it to a particular phase by time-shifting the folded light-curve.
Since the data are noisy it will not be practical to just finding
the maximum/minimum value of the magnitude. On the contrary, we
must find a statistical measure of the signal.

Our method is based on a clustering technique that
divides the data (here data refers to a single light-curve) 
into clusters (cluster here means a subset of observations within a light-curve) 
based on the magnitude and then finds the cluster with the maximum average.

To find the clusters we required that both the density within
the clusters and the separation between clusters should be maximum.
In other words we want the clusters to be as compact as possible and be as separated from
other clusters as possible.

We measure the cluster compactness or inter-cluster measure of two clusters as:
\begin{equation}
  S_{{\rm inter}} \equiv  \sum_{t_i \in C1}\left( t_i-\bar{t}_{C1} \right)^2 +  \sum_{t_i \in C2}\left( t_i-\bar{t}_{C2} \right)^2 \; ,
\end{equation}
where $C1$ and $C2$ denote the clusters, $t_i$'s are the times of observations in the particular cluster and
$\bar{t}_{C1}$ is the average time in cluster $C1$. We also define the intra-distance between the two clusters as
\begin{equation}
  S_{{\rm intra}} \equiv  \frac{ \left| \bar{t}_{C1} -  \bar{t}_{C2} \right| }
            { \sqrt{  \frac{\sigma_{C1}^2 }{ N_{C1}} +\frac{\sigma_{C2}^2 }{ N_{C2}}}} \,\, ,
\end{equation}
where $N_{C1}$ is the number of points in cluster $C1$ and
\begin{equation}
  \sigma_{C1}^2 = \frac{1}{ (N_{C1} -1 )} \sum_{t_i \in C1}\left( t_i-\bar{t}_{C1} \right)^2  \: .
\end{equation}

We define the following measure which by minimizing  gives us
a measure of goodness of clustering,
\begin{equation}
  S \equiv \frac{S_{{\rm inter}}}{ S_{{\rm intra}}} \: .
\label{eq:clustermes}
\end{equation}
The actual algorithm is described below:
\begin{itemize}
\item For each light-curve we select the highest/lowest 10\%
magnitude data points. \item We divide the data in two clusters as
$ t_{c1} \in \{ t_1,t_2, \ldots, t_s \} $ and $ t_{c2} \in \{
t_{s+1}, t_{s+2}, \ldots, t_N \} $ where $s$ is the index of the
separator. \item For each $s = \{1 \ldots N \}$ we calculate the
goodness of clustering using Eq.~\ref{eq:clustermes}. If $S$ is
minimum within the range $1<s<N$ we keep the division of data into
two clusters. We repeat this process in the sub-clusters until no
more clustering is favorable \footnote{Since the data are in a
dimensional space it is guaranteed that points in the same cluster
are sequential. Therefore a separation at a given iteration  cannot
alter the clustering measure of the previous iteration}. \item After
the clustering is done we calculate the mean magnitude and mean
time in each cluster. We select the cluster of the highest mean
magnitude. \item We translate time such as the mean time of the
selected cluster is always at the same predefined time.
\end{itemize}

\begin{figure}
\begin{minipage}{7cm}
   \includegraphics[width=7cm]{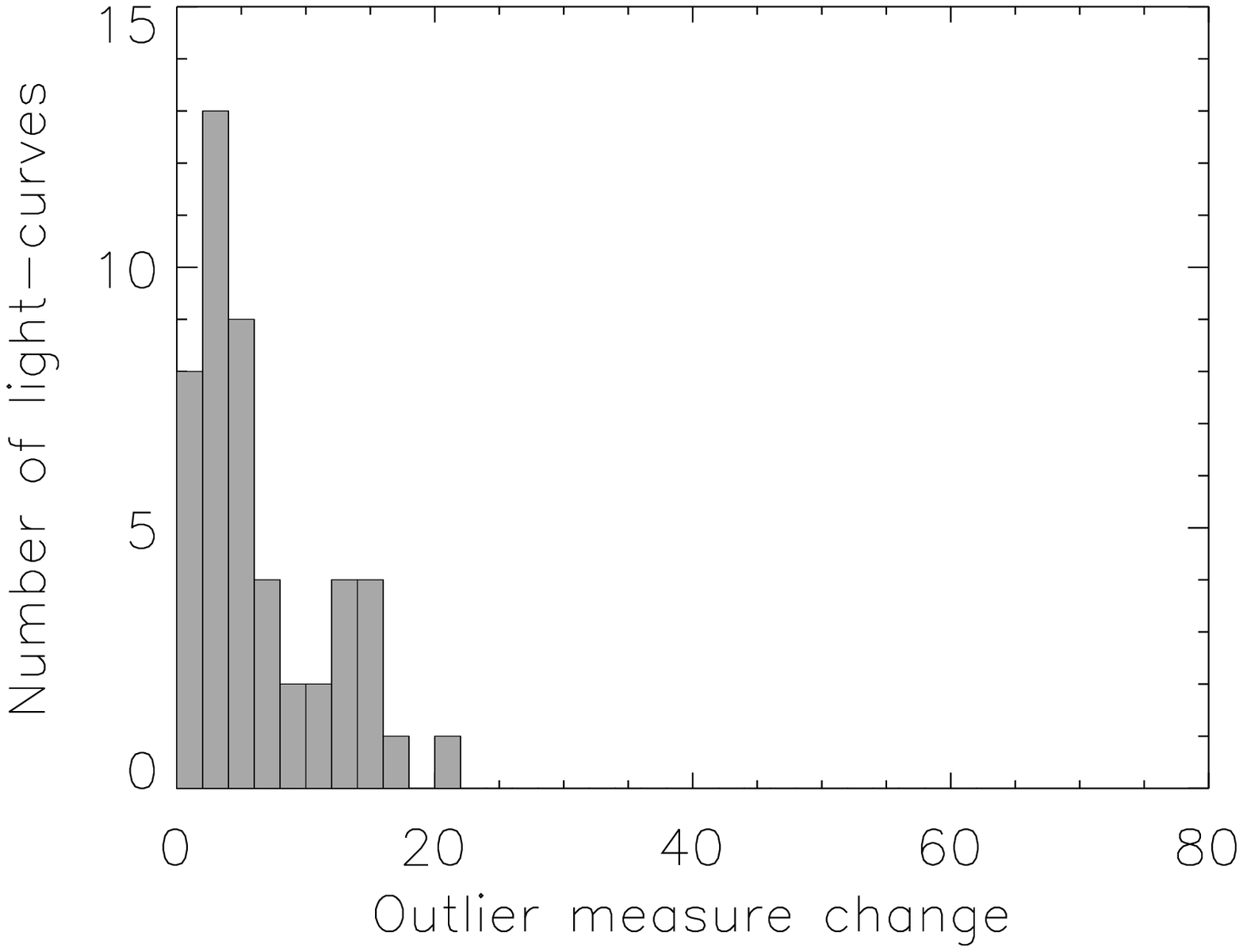}
  \caption{ {\small Histogram of the outlier measure difference between the full method and the approximate
method for 500 EB's from OGLE EB catalog. Only the 10\% with the lowest average correlation were used. }}
  \label{fig:rankdiff500eb}
\end{minipage}
\hspace{.32cm}
\begin{minipage}{7cm}
    \includegraphics[width=7cm]{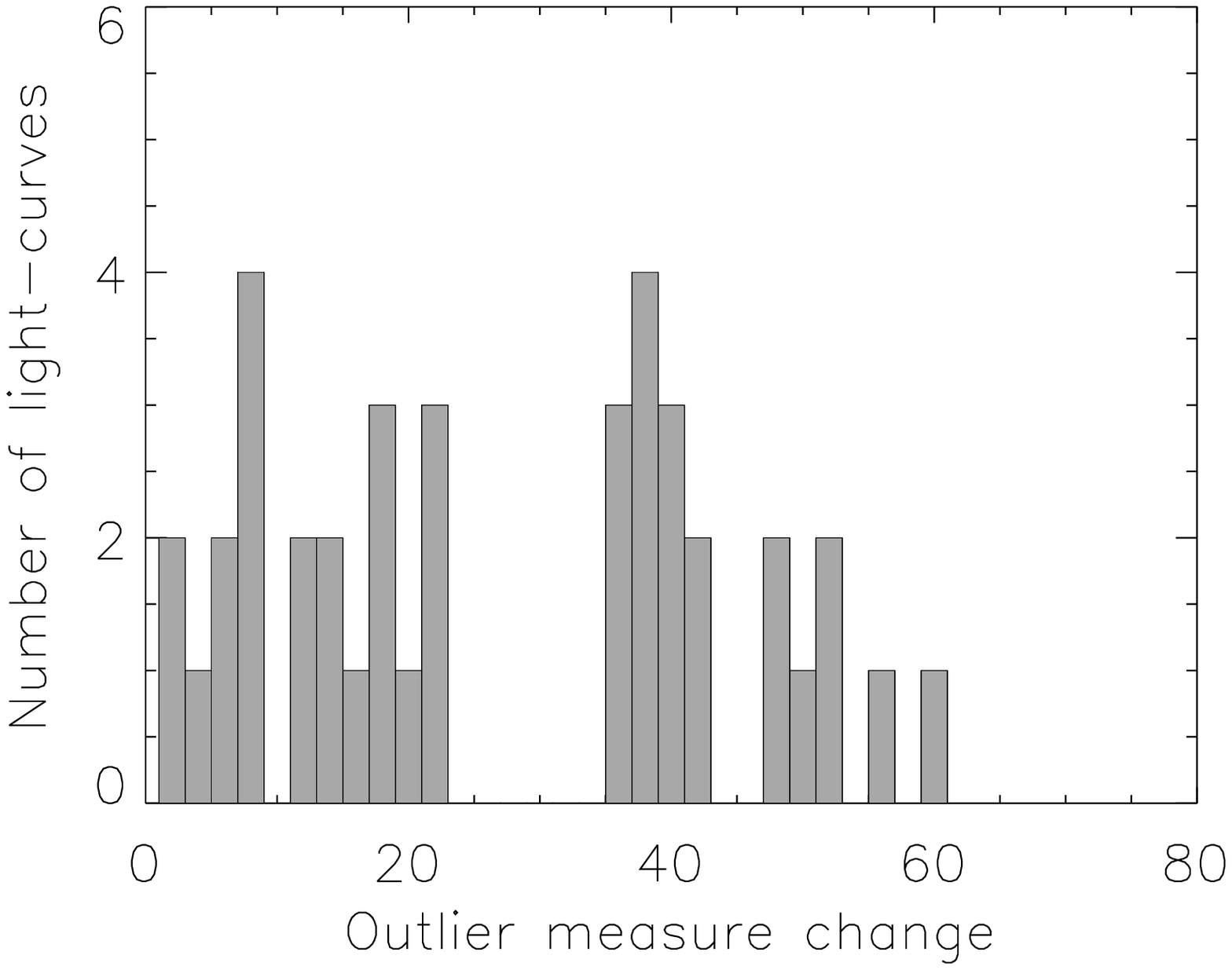}
    \caption{ {\small Histogram of the outlier measure difference between the full method and the approximate
method for 1000 \rrs~ from the OGLE \rr~ catalog. Only the 10\% with the lowest average correlation were used. }}
  \label{fig:rankdiff1000rrl}
\end{minipage}
\end{figure}

By phasing every light-curve to a universal phase the method of
``correlation-to-the-average'' can be applied assuming that the
observational errors are  incorporated in the running average
method. However the method is an approximation since it does not
guarantee that the correlation between two light-curves is
maximum. Nevertheless for most light-curves where a
maximum/minimum signal is well defined this method should give us
very similar results to the full method. We have tested this
method on two sets; 500 light-curves of OGLE Eclipsing Binary
stars (EBs) and 1000 light-curves of OGLE \rr stars. 
Fig.~\ref{fig:rankdiff500eb} and Fig.~\ref{fig:rankdiff1000rrl}
show the runs on these two sets. In each figure we show a
histogram of the the rank differences between the full method and
the approximation described in this section for the bottom 10\% of
the light-curves.   EBs do have a much better defined minimum, so
the approximation performs very well (most light-curves are
ranked with $\pm10$ of the original rank), whereas the case of
\rr's the approximation is not performing as well.

\subsection{Outlier analysis within subsets}
\label{sec:subsets} Another alternative approach which avoids the
drawbacks of the method described above is based on a simple
statistical argument. If a light-curve is an outlier in the whole
set it will be an outlier in a large subset of the whole set. We
could then in principle divide the whole set into large subsets and
perform the analysis on each subset.
If the subsets are randomly selected and the number
of light-curves is large enough the outlier measure from each
subset can be put together and hence we can rank all  light-curves
as if they were in a single set.

Since each subset must be
a substantial fraction of the full set ($\geq 10$\%) the overall
performance gain is about a factor of ten at best. In the case of large
sets this method will not scale very favorably but
it is an ``exact'' method and it is very easily parallelizable.

We have applied this method to
16,020 of the \rrs~ from the MACHO survey (see Sec.~\ref{sec:results}).

%-----------------------------------------------------------
%
%          Figure performance
%
%-----------------------------------------------------------

%-----------------------------------------------------------
%
%          Figure rankdiff
%
%-----------------------------------------------------------

%============================================================================
%============================================================================
%         RESULTS
%============================================================================
%============================================================================
\section{Results}
\label{sec:results}

% ----------------------------------------------------------------
%         data
% ----------------------------------------------------------------
%\subsection{Data}
We tested the validity of our method on various periodic star
catalogs, both published and unpublished, compiled by the MACHO
collaboration \citep{alcock00}
\footnote{http://www.macho.mcmaster.ca/} and by the OGLE
collaboration \citep{udalski97}
\footnote{http://sirius.astrouw.edu.pl/\~{}ogle/}.
\par
Both the MACHO and the OGLE projects were microlensing surveys
devoted to finding gravitational microlensing events in the halo
of the Milky Way by background stars in the Large and Small
Magellanic Cloud (LMC and SMC) and the bulge of the Milky Way.
These surveys also produced large catalogs of variable stars:
details on the MACHO variable star research can be found in
\cite{machovar1,machovar2,machovar3,machovar4,cook95}. OGLE
variable star catalogs found during part II of the project
(OGLE-II), \citep{udalski97} with accompanying papers,
can be found on the group website \citep{Soszynski03,udalski99,udalski99b,wyr03}.
\par
The variable stars considered were Eclipsing Binary stars (EBs) of
which catalogs were published by MACHO
\citep{faccioli05,lacy97} and OGLE \citep{wyr03}, \rr~
and Cepheids from OGLE \citep{udalski99,wyr03} and 
unpublished MACHO collections that were compiled at Lawrence Livermore
National Laboratory (LLNL) by Kem Cook, Doug Welch and Gabe
Prochter. These lists have been generated from the MACHO database
by appropriate cuts in the period-luminosity diagram.  This 
is only a first step in producing a catalog and thus the resulting lists 
are expected to be contaminated.
\par
MACHO observations were taken in two non standard band passes:
MACHO ``blue", hereafter indicated as $V_{\mathrm{MACHO}}$, with
a bandpass of 440-590nm and MACHO ``red", hereafter indicated as
$R_{\mathrm{MACHO}}$, with a bandpass of 590-780nm;
transformations to standard Johnson $V$ and Cousins $R$ bands are
described in detail in \citep{alcock99}. \footnote{Transformation
to standard magnitudes is given by (Kem Cook, private
communication):
\begin{eqnarray}\label{eq:machocal}
V&=&V_{\mathrm{MACHO}}+24.22-0.1804(V_{\mathrm{MACHO}}-R_{\mathrm{MACHO}})
\nonumber \\
R&=&R_{\mathrm{MACHO}}+23.98+0.1825(V_{\mathrm{MACHO}}-R_{\mathrm{MACHO}})
\nonumber
\end{eqnarray}
}
\par
The average number of observations in both bands is several
hundreds, with the center of the LMC being observed more
frequently than the periphery.
\par

MACHO periods were found by applying the {\it Supersmoother} algorithm
\citep{reimann94}, first published by \cite{fri84}. The algorithm
folds the light-curve around different trial periods and selects
the one that gives the smoothest folded light-curve. Periods were
found for the red and the blue band separately, and usually agree
with each other to better than 1\%. The algorithm may fail though,
usually determining a period for one color band that is a multiple
of the period found for the other band.  In these cases
the light-curve with the incorrect period will often be
flagged as an outlier; hence the program can be useful in finding
wrong periods in a large data set of variable stars (see MACHO
Cepheids and \rrs~ below).
\par
OGLE observations were taken in the $B$, $V$ and $I$ bands and
reduced via Difference Image Analysis (DIA) \citep{zebrun01a}; a
catalog of variable stars for the Magellanic Clouds was thus
produced \citep{zebrun01b} and from it a sample of 2580 EBs was
selected \citep{wyr03}; we used only $I$ band, DIA reduced
observations in our analysis, since the number of observations in
this band was much higher (on the order of $\approx$ 200-300)
compared to $V$ and $B$. \vspace{.1cm}
\par
The main features of the MACHO and OGLE variable star datasets are summarized in
Table~\ref{tab:catalogues}.
\begin{table}
\caption{Main features of the catalogs used}
\label{tab:catalogues}
\begin{tabular}{lcc}
\hline  \hline
Type of Variable star & Found by & Number of Stars \\
\hline
%%\tablecolumns{3}
%%\tablewidth{0pc}
%%\tablecaption{Main features of the catalogs used in the testing
%%phase}
%%\tablehead{
%%\colhead{Type of Variable star} &
%%\colhead{Found by:} &
%%\colhead{Number of Stars}
%%}
%%\startdata
Cepheids & MACHO &  3177$^{\dagger}$ \\
Cepheids & OGLE-II & 1329$^{\dagger}$ \\
RRLyr\ae & MACHO &  16020$^{\dagger}$  \\
RRLyr\ae & OGLE-II & 5327$^{\dagger}$ \\
EBs & MACHO & 6064$^{\dagger  \ddagger}   $  \\
EBs & OGLE-II & 2580$^{\dagger}$ \\
\hline \multicolumn{3}{l}{$^\ddagger$ EBs from both LMC and SMC
were included.} \\
 \multicolumn{3}{l}{$^\dagger$ Only light-curves with at least 100 observations were included.} \\ \hline
%%\enddata
%%\tablenotetext{a}{Of which 5000 randomly chosen objects were used in
%%the trials.}
\end{tabular}
\end{table}

% -----------------------------------
%
% -----------------------------------
%\subsection{Testing the method}
%
%
\vspace{.1cm} Results of these runs are presented in the following
way: For each of the collections listed in Table 1  there are three figures and one
table. The first figure shows the histogram of the outlier
measure. The second figure shows the centroid light-curve as
defined in Eq.~\ref{eq:centroid}. The next 9-panel figure presents
the lowest nine light-curves, i.e. our outliers. Each panel is
labeled according to its position in the figure; from A1 to C3.
Following that there is a table which summarizes the properties of
these outliers including our interpretation. These interpretations
were formed after further investigation including cross
correlation with other surveys, position in the HR diagram,
spectral types where available, etc.

\vspace{.3cm}\par \noindent
 \underline{Cepheids:}
%The term Cepheid was at one time applied to any continuously
%varying star with regular light-curve and a period less than 35
%days. It is now recognized that this classification is diverse,
%containing stars in different mass ranges and evolutionary states.
%Stars with period of one day or less are now mainly classified as
%\rr~ variables and RV Tauri stars are also treated differently. The
%remaining stars are called $\delta$Cephei variables, Type I
%Cepheids, classical Cepheids or simply Cepheids. 
Cepheids are periodic variables with periods ranging from about 1 day to about
50 days (with few extreme examples of 200 days) and which lie between 
the main sequence and the giant  stars. 
. Detailed
characteristics of their light-curves varied depending on the
period (Hertzprung progression). More details about Cepheids and other
variable stars in general can be found in
\cite{Petit87} and \cite{Sterken96}.

\vspace{.1cm}
\par
\begin{figure*}
\begin{minipage}{7cm}
   \includegraphics[width=6cm]{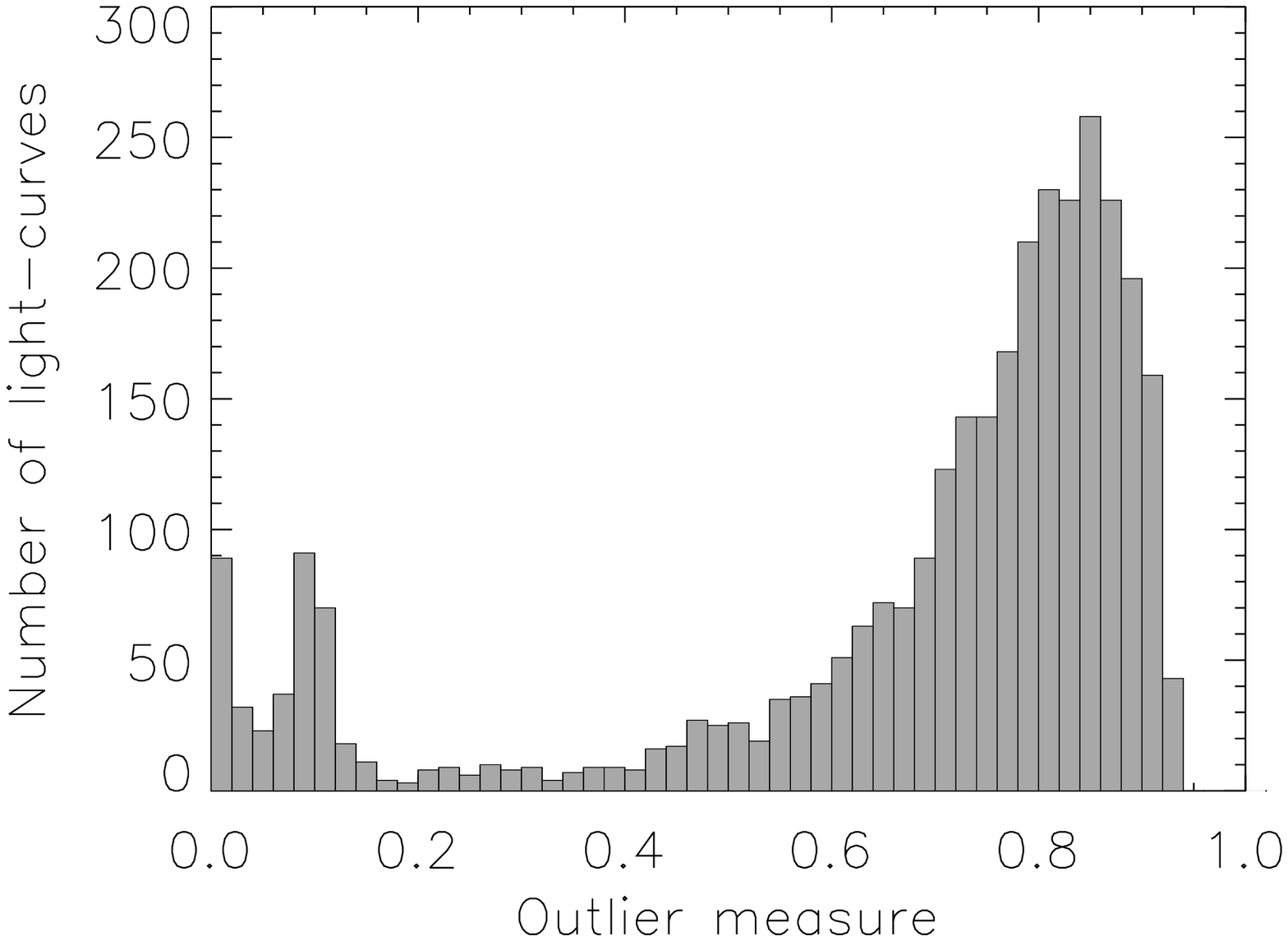}
  \caption{ {\small Histogram of the outlier measure for 3297 {\bf Cepheids}  in the
{\bf MACHO} sample.}}
  \label{fig:MACHO_ceph_hist}
\end{minipage}
\hspace{1cm}
\begin{minipage}{7cm}
    \includegraphics[width=6cm]{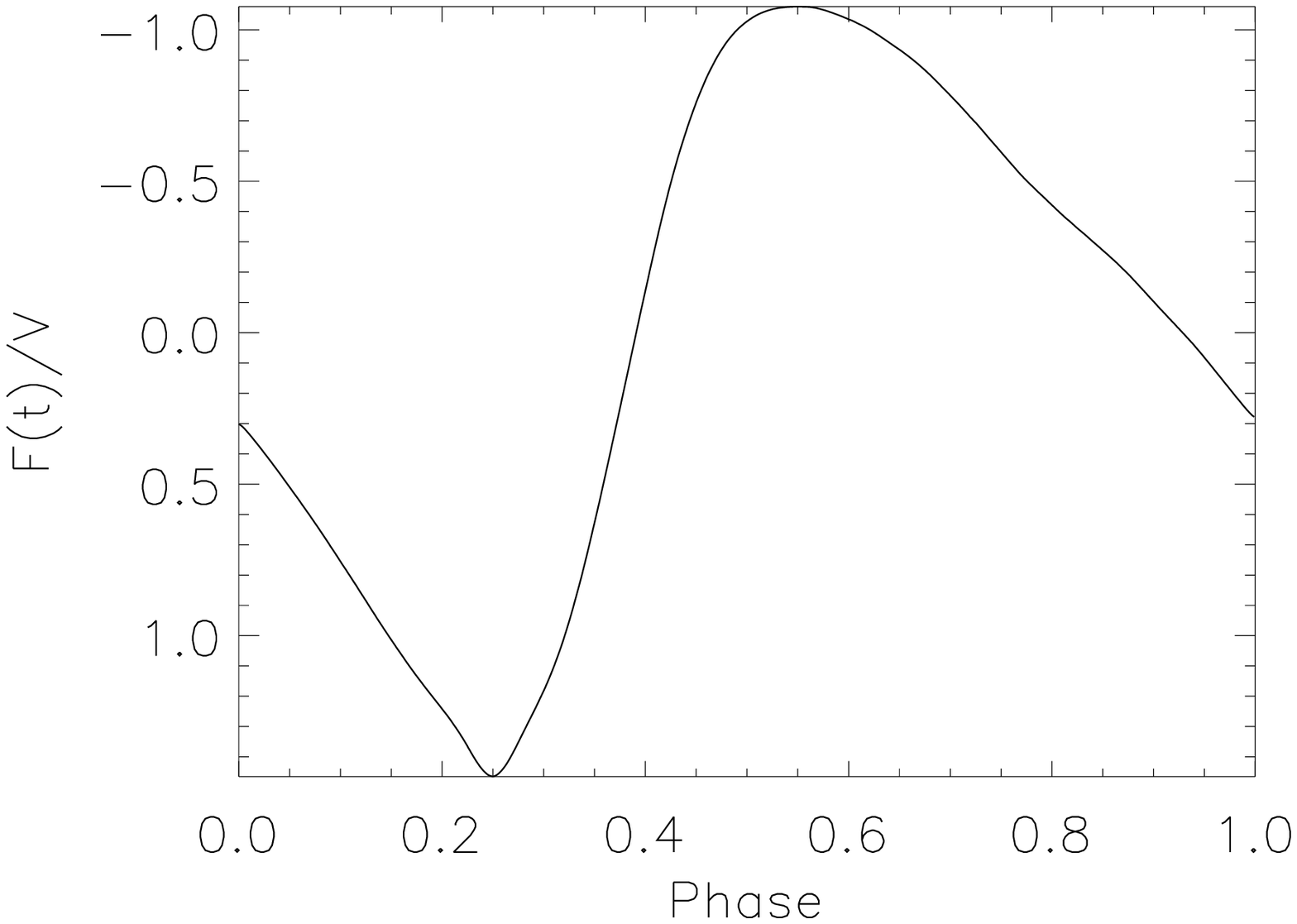}
    \caption{ {\small Centroid light-curve for  3297 {\bf Cepheids}   in the
{\bf MACHO} sample.}}
\end{minipage}
\end{figure*}

\begin{figure*}
 \includegraphics[width=18cm]{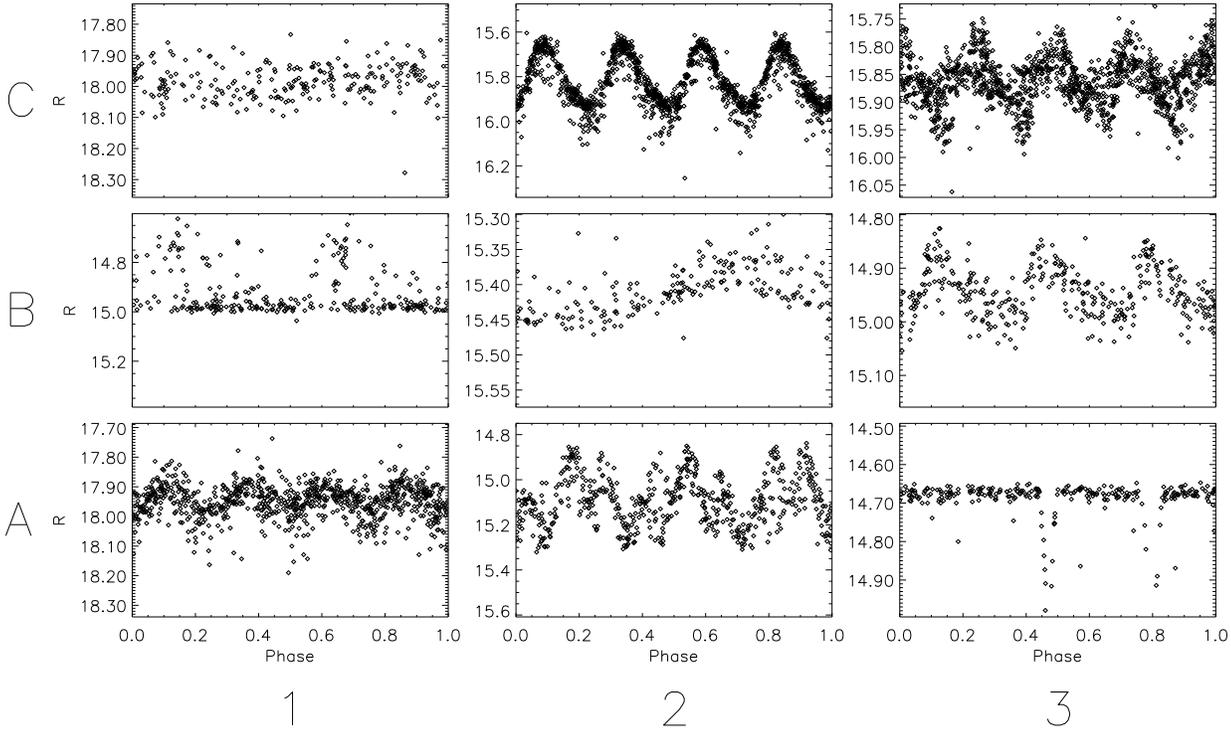}
\caption{Light-curves with the lowest  measure of similarity from
the {\bf MACHO Cepheid} dataset. Only the RED band is shown.}
\label{fig:MACHO_ceph_outliers}
\end{figure*}

 \noindent The MACHO Cepheid dataset contains a small number of light-curves where
the folded period is an integer multiple of the ``correct''
period. This can be seen in Fig.~\ref{fig:MACHO_ceph_outliers}-A1, A2, B3, C2, C3 .
Also there is a second bump in the histogram of average correlations (Fig.~\ref{fig:MACHO_ceph_hist})
at about 0.1. These light-curves are mostly light-curves folded with integer multiple
period of the "true" period. Notwithstanding, the light-curve shown in A3
in the same figure is clearly an EB and not a Cepheid.  B1 is
evidently a periodic light-curve (apparent from the distinct
pattern in the folded light-curve) but the shape in both R and V
bands (only R shown) does not match that of a Cepheid (or all
subtypes). Further investigation (e.g. spectral type) is needed
to determine the type of variable. Note our goal in this work is
to identify the outliers and thus demonstrate that this method can
lead us to the few interesting cases. It is not our intention to
do an in depth investigation for each unidentified light-curve only
to point out the obvious misclassification's and interesting
cases. Light-curve shown in C1 does not look periodic or variable for that matter
in both bands thus we classify it as ``likely not periodic''
star.

\vspace{.1cm}
\par
\begin{figure*}
\begin{minipage}{7cm}
   \includegraphics[width=6cm]{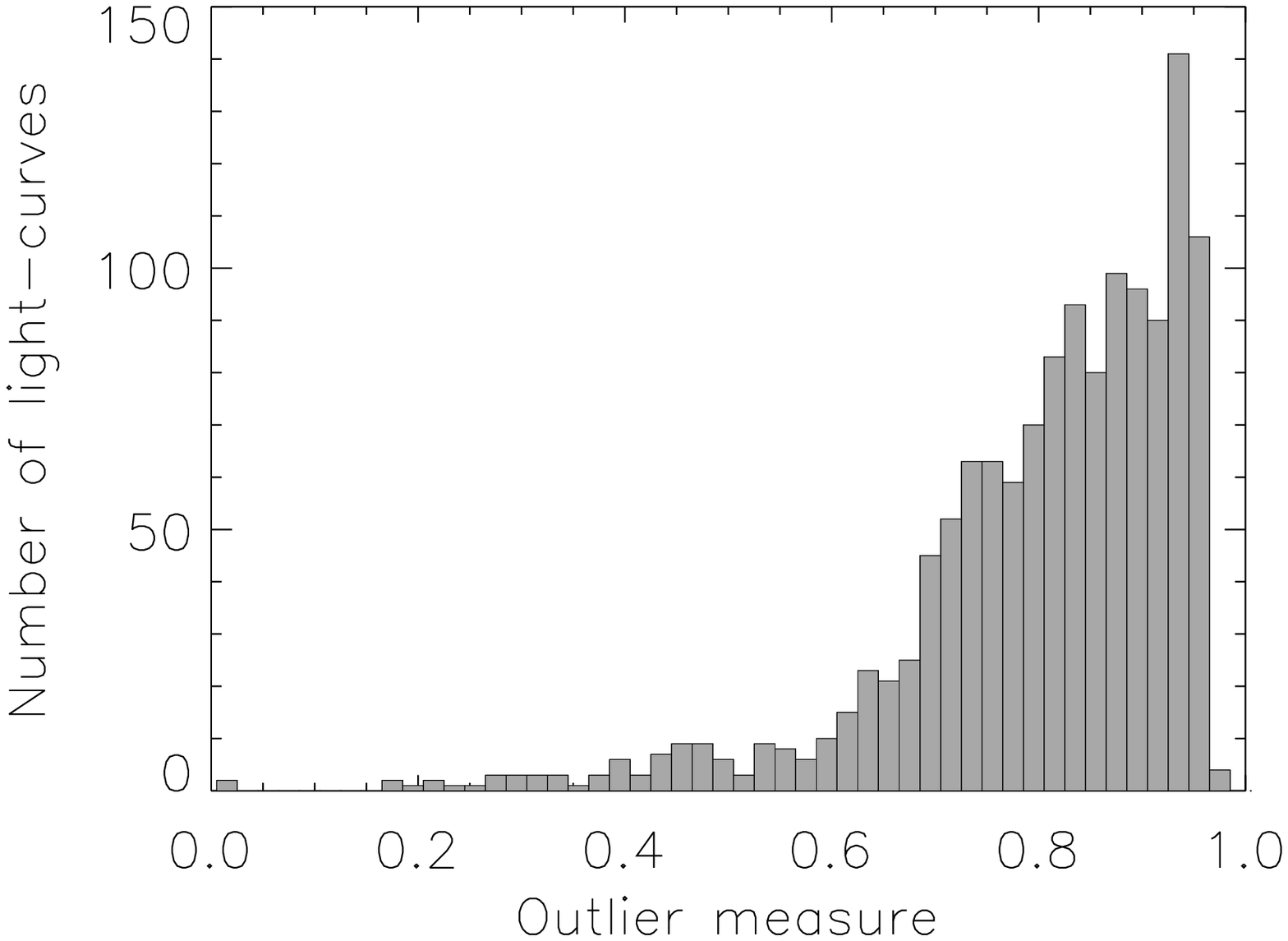}
  \caption{ {\small Histogram of the outlier measure for 3297 {\bf Cepheids}   in the
  \label{fig:OGLE_ceph_hist}
{\bf OGLE} sample.}}
\end{minipage}
\hspace{1cm}
\begin{minipage}{7cm}
    \includegraphics[width=6cm]{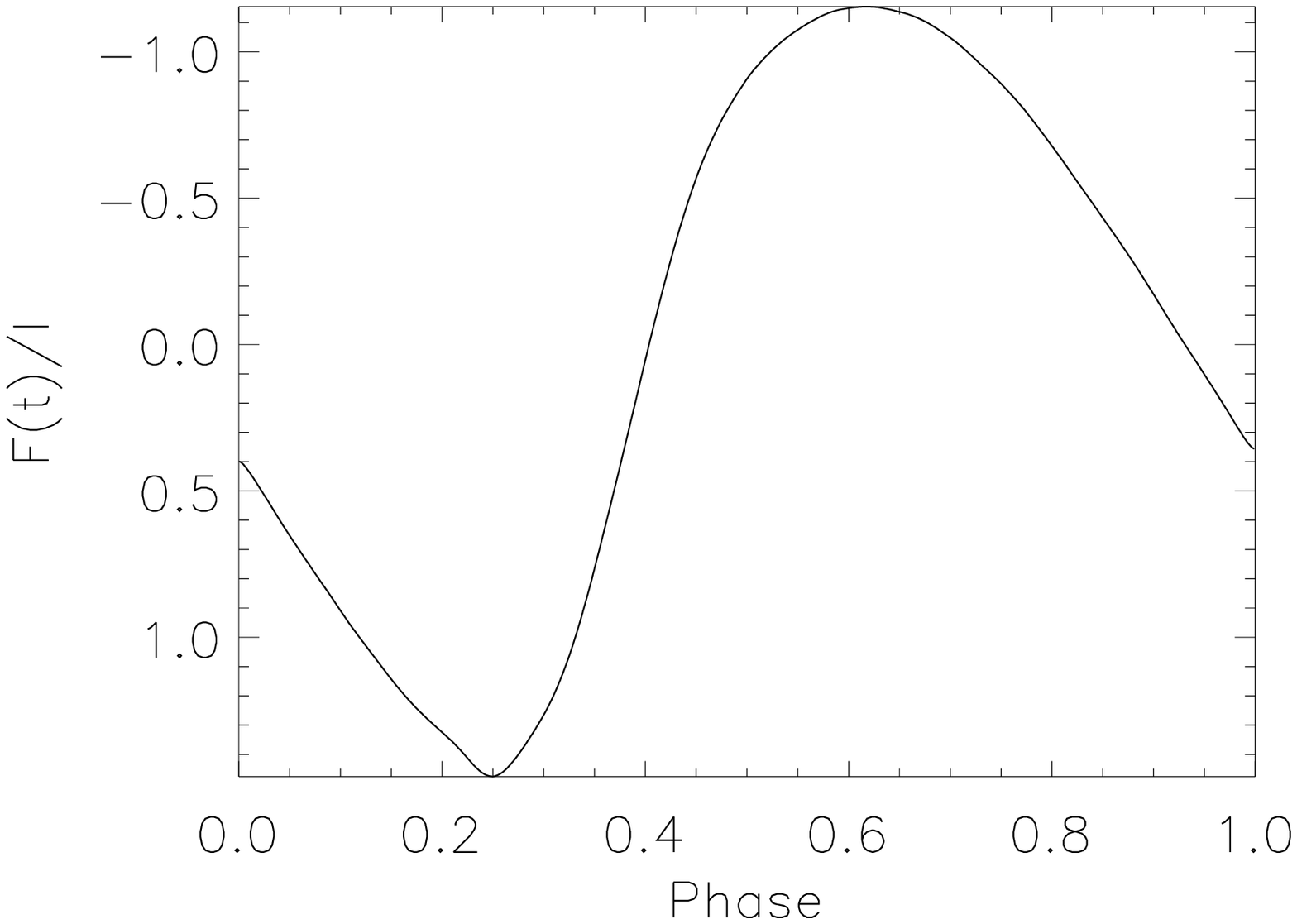}
    \caption{ {\small Histogram of the outlier measure for 3297 {\bf Cepheids}   in the
{\bf OGLE} sample.}}
\end{minipage}
\end{figure*}

\begin{figure*}
 \includegraphics[width=18cm]{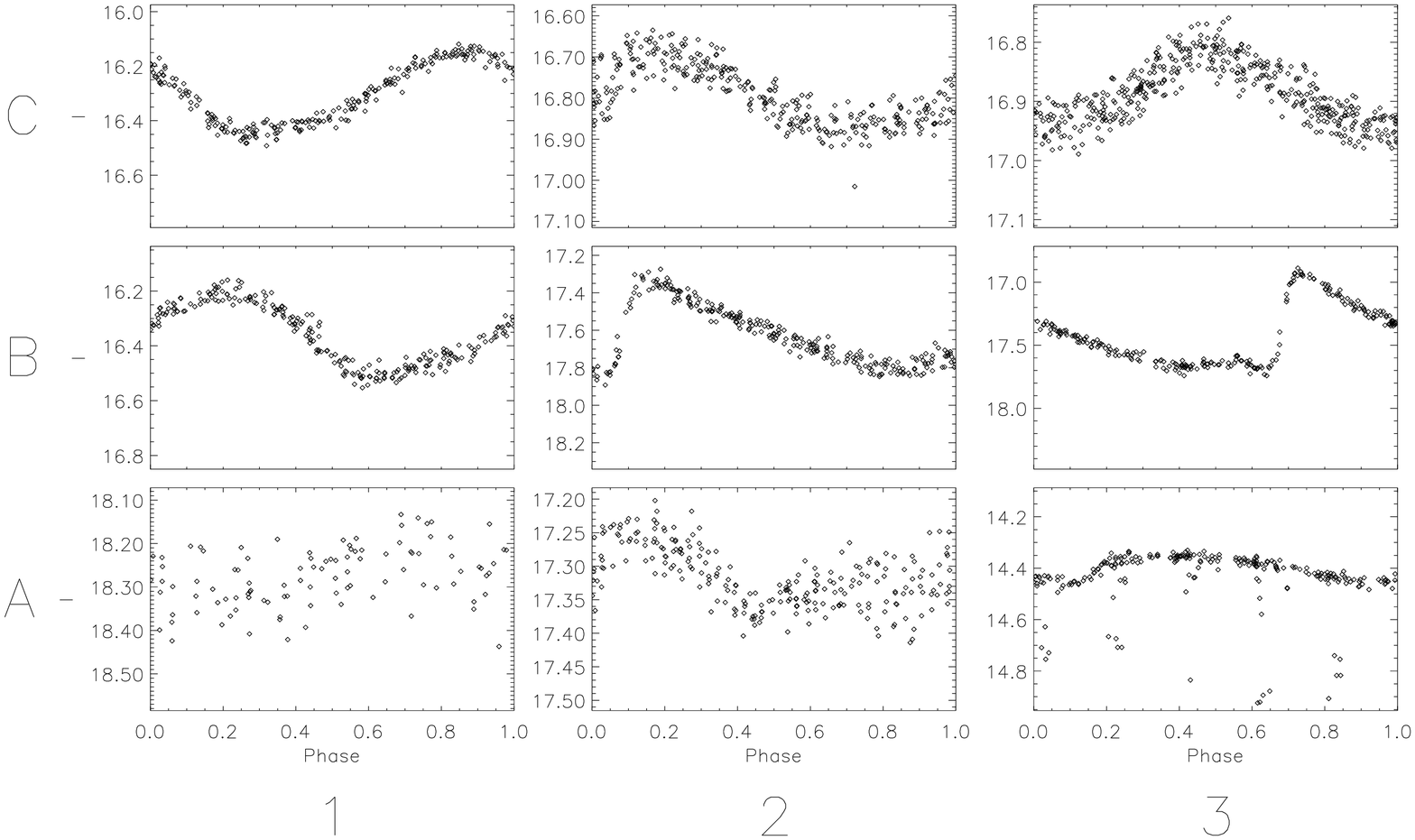}
\caption{Light-curves with the lowest  measure of similarity from
the {\bf OGLE} {\bf Cepheid} catalog. Only the  {\bf
OGLE} I band is shown.} \label{fig:OGLE_ceph_outliers}
\end{figure*}

 \noindent The OGLE Cepheid catalog \citep{udalski99,wyr03} has few 
true outliers. Only three
interesting cases did make it into our list (see
Fig.~\ref{fig:OGLE_ceph_outliers} and
Fig.~\ref{fig:OGLE_ceph_hist}). From the histogram in
Fig.~\ref{fig:OGLE_ceph_hist} we see that there is no second bump
but three light-curves are clearly on the lowest bin. A1 is vaguely
a periodic light-curve but there are not enough data and
they are too noisy. Even if we agree to the  periodicity
we find that the asymmetry is atypical of Cepheids of all types, with
slow rise and fast decline.
Similarly A2 exhibits a clear periodic signal but wrong asymmetry.
Light-curve A3 is interesting. The overall shape, period and color are
consistent with a Cepheid. The extra regularly spaced spikes are  too
regular in folded space to be ignored. The possibility to be
an EB with Cepheid variable is highly unlikely since the periods are
synchronized (1:5) which suggest some fundamental dynamical
process. A more careful study is needed to understand the
physical process underlying this light-curve.
The rest of the outliers have much higher average
outlier measure and they are only shown here for consistency (9
light-curves per catalog).

\vspace{.1cm}\par \noindent \underline{\rrs:} \rrs~ come in many
different types but most predominately in two subclasses. The RRAB
which is the majority of them and RRC. These pulsating stars have
very well defined period (0.5-0.3 days). They are usually
asymmetric however a subclass RRS does have a sinusoidal shape. 
It is usually hard to distinguish them from Cepheids just from the
characteristics of the shape of the light-curves. More details 
about \rr stars can be found in \cite{Petit87} and \cite{Sterken96}.

\par
\begin{figure*}
\begin{minipage}{7cm}
   \includegraphics[width=6cm]{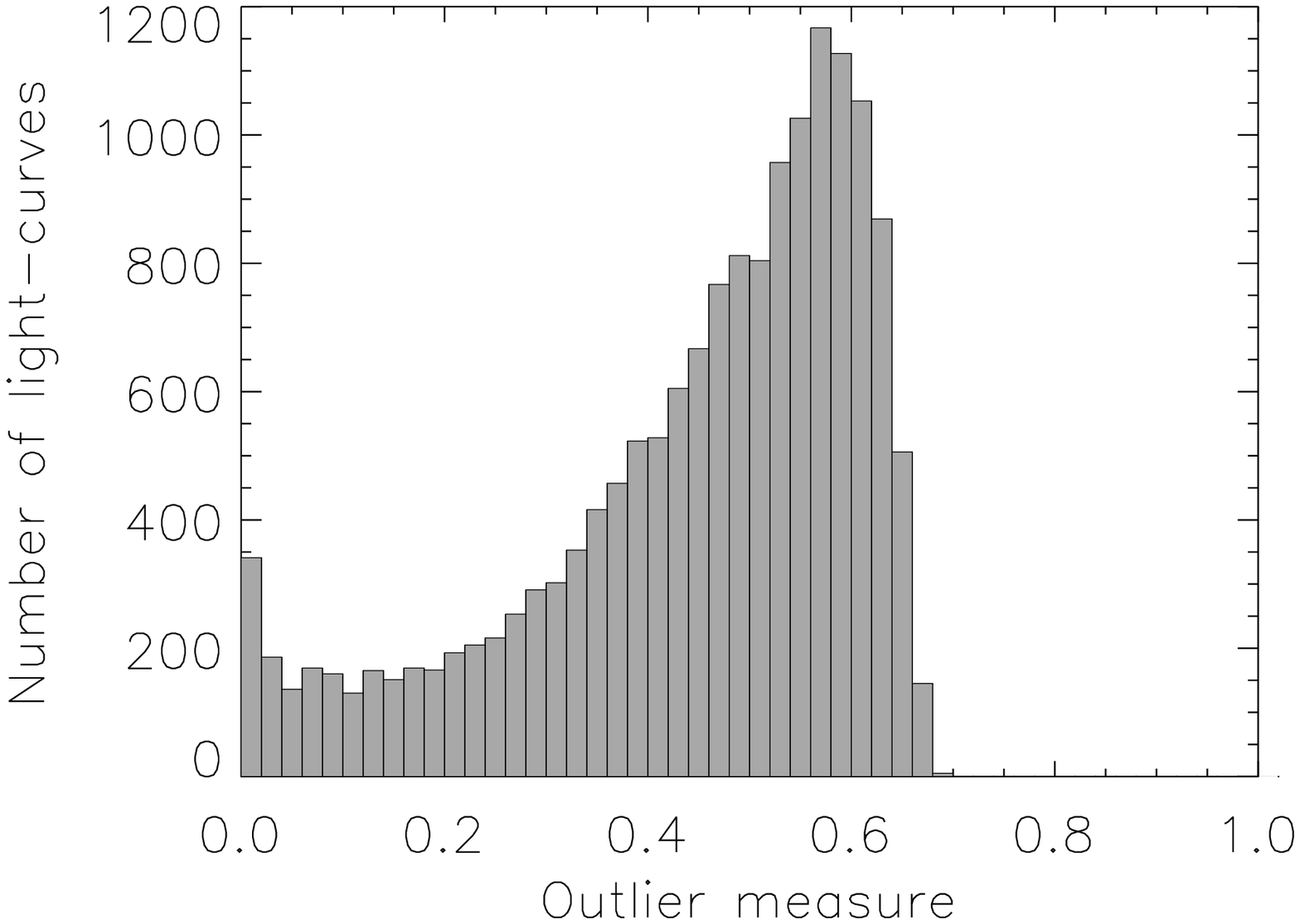}
  \caption{ {\small Histogram of the outlier measure for 16080 {\bf RRL}  in the
{\bf MACHO} sample.}}
  \label{fig:MACHO_rrl_hist}
\end{minipage}
\hspace{1cm}
\begin{minipage}{7cm}
    \includegraphics[width=6cm]{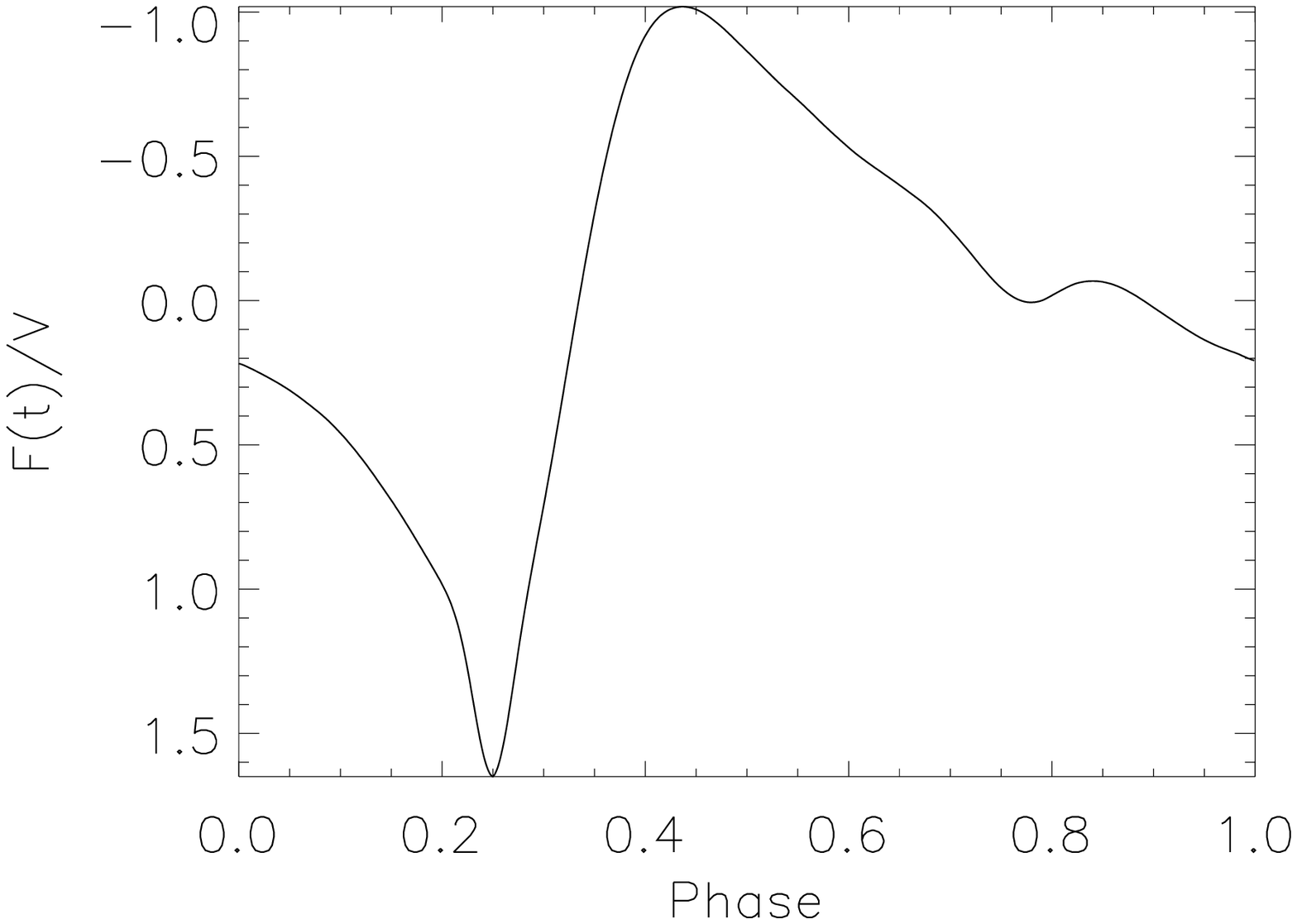}
    \caption{ {\small Centroid light-curve  for 16080 {\bf RRL}  in the
{\bf MACHO} sample.}}
\end{minipage}
\end{figure*}

\begin{figure*}
\begin{minipage}{7cm}
   \includegraphics[width=6cm]{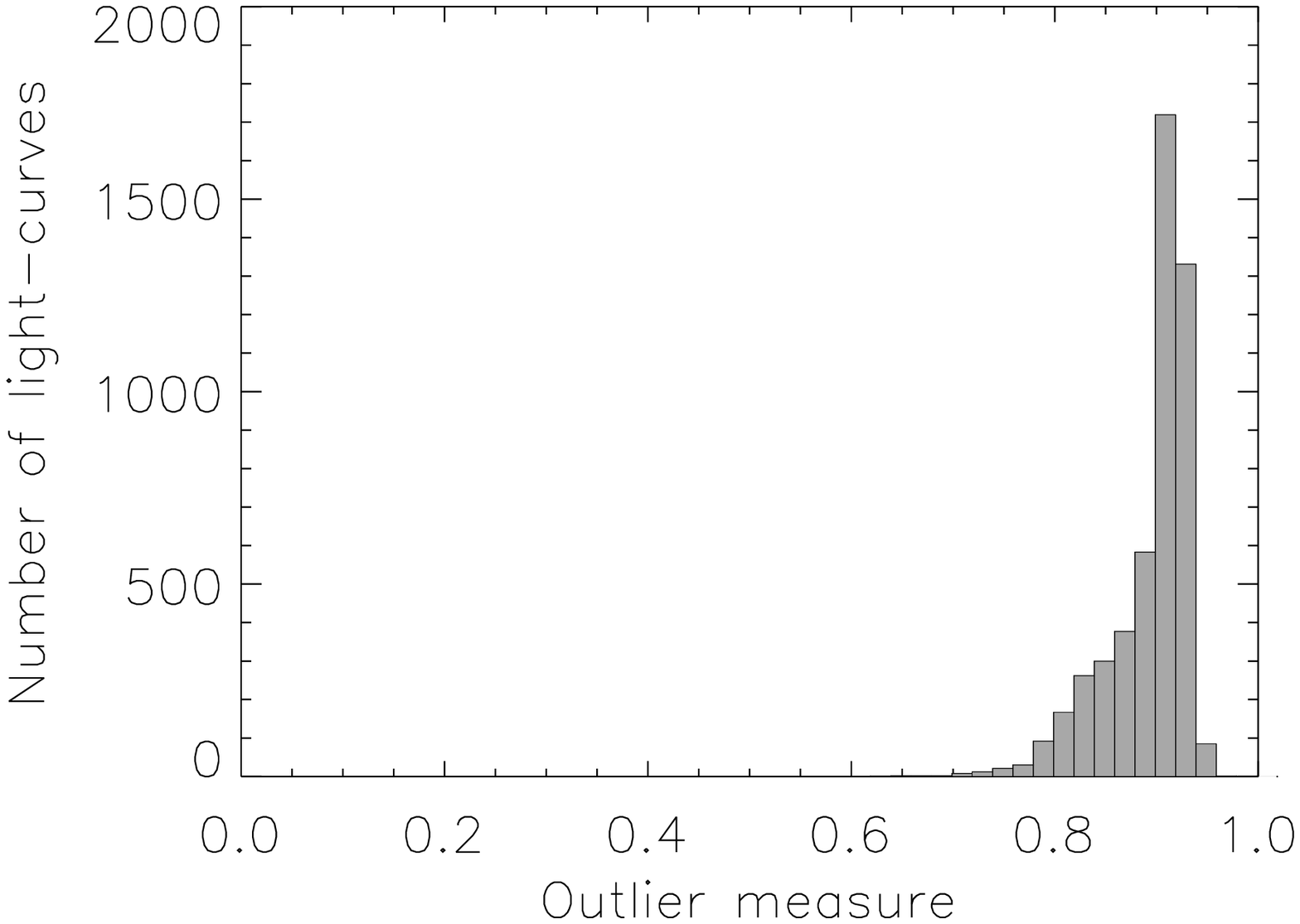}
  \caption{ {\small Histogram of the outlier measure for  {\bf RRL}  in the
{\bf OGLE} sample.}}
  \label{fig:OGLE_rrl_hist}
\end{minipage}
\hspace{1cm}
\begin{minipage}{7cm}
    \includegraphics[width=6cm]{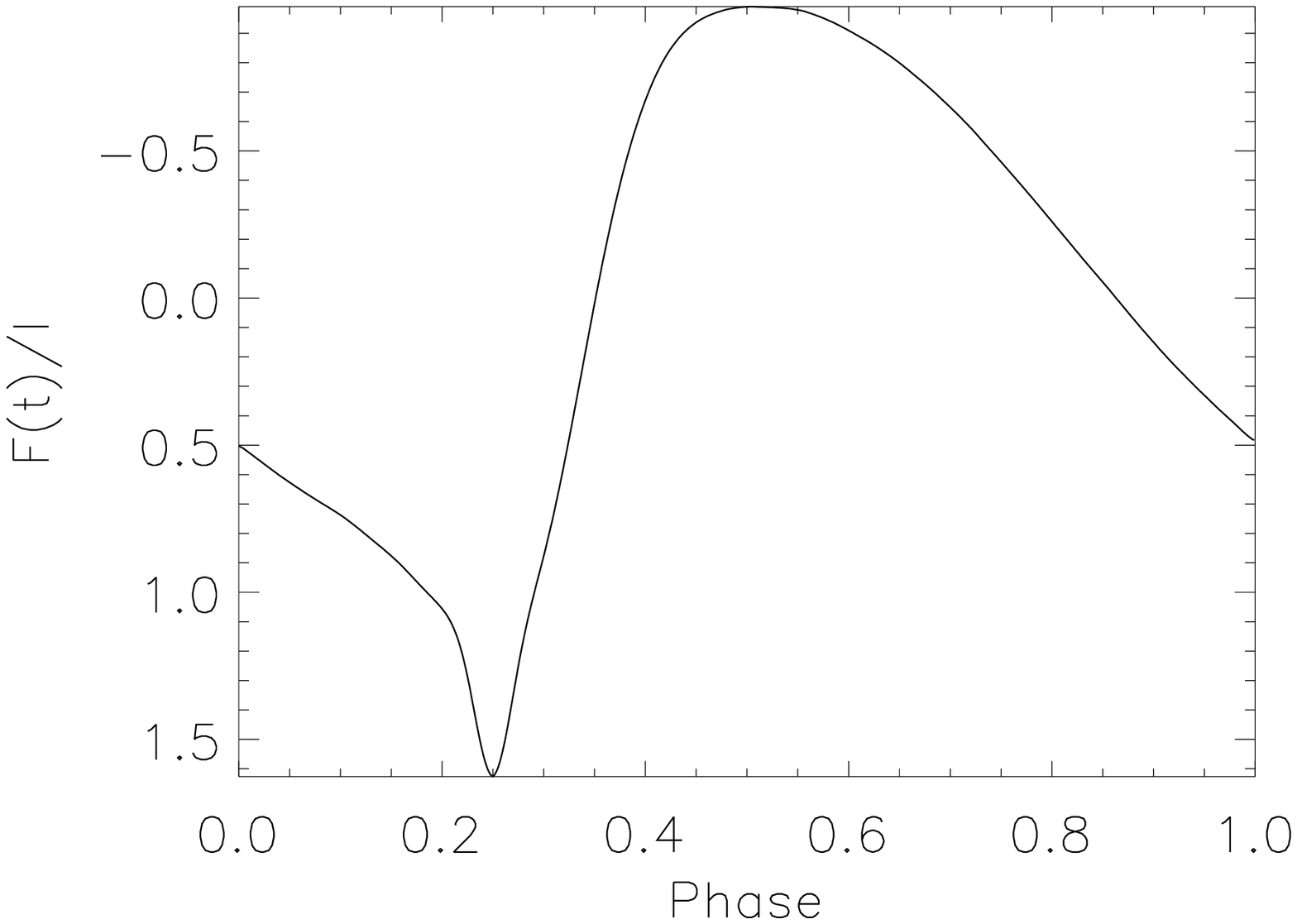}
    \caption{ {\small Centroid light-curve  for 5327 {\bf RRLs}  in the
{\bf OGLE} sample.}}
\end{minipage}
\end{figure*}

\noindent
The published
OGLE catalog  \citep{udalski99} is ``cleaner'' (does not contain the wrong types or
wrongly stated period of variables) than the unpublished MACHO
collection. This can be seen from Fig.~\ref{fig:MACHO_rrl_hist} and
\ref{fig:OGLE_rrl_hist} where it is clear that the correlation
distribution of the MACHO dataset is centered closer to zero than
the distribution of the OGLE catalog (this is due to contamination 
of the MACHO dataset with other variable stats). 

\vspace{.1cm}
\par
\begin{figure*}
 \includegraphics[width=16cm]{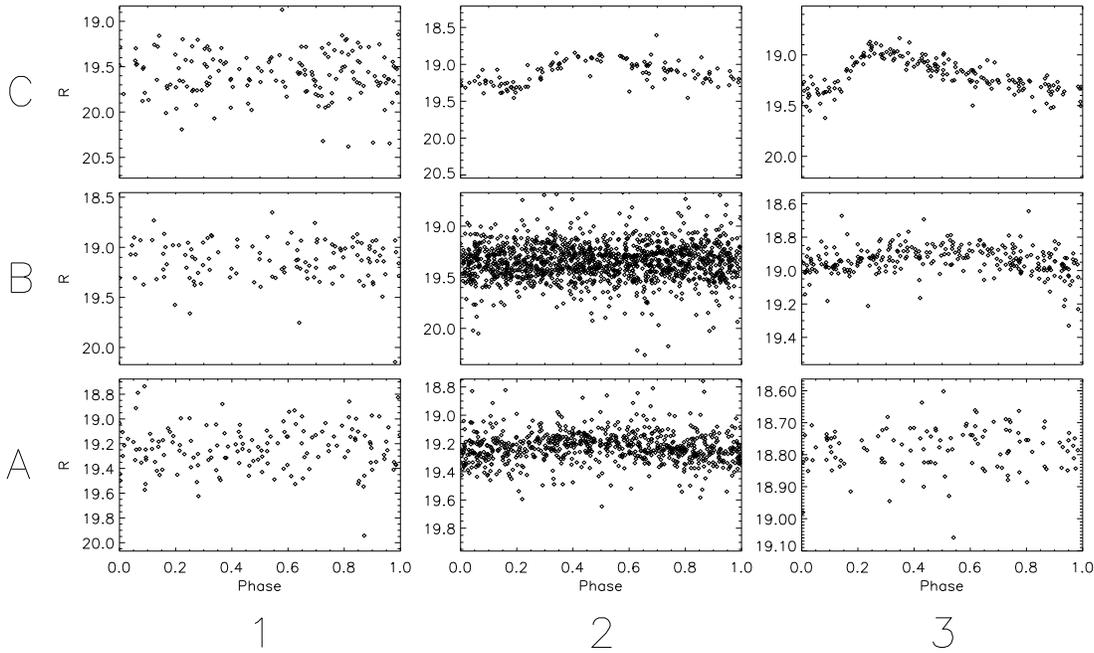}
\caption{Light-curves with the lowest  measure of similarity from
the {\bf MACHO RRL} dataset.  Only the R band is shown.}
\label{fig:MACHO_rrl_outliers}
\end{figure*}

 \noindent As in the case of Cepheids
the \rr~  MACHO dataset contains light-curves that are either folded
 using a multiple of the true period or folded simply with the wrong period
in one of the two bands and thus appear to be outliers.
Nevertheless some of the light-curves were most likely
misclassified as \rrs. Light-curves A2 and A3 in
Fig.~\ref{fig:MACHO_rrl_outliers} have periods of 0.98 and 0.53
days which are too large to belong to RRC group. Such periods can be
from the RRAB group but the shape, amplitude and symmetry of the
light-curves indicates a non periodic light-curve; hence we ruled
them as possibly misclassified. A1 was identified as the outlier of greatest degree.
However when we looked at the V-band light-curve it had
the characteristics (period, amplitude etc) of a RRC.
Light-curves B1, B2 and C1 were simply folded with the wrong period
in the red band. Looking in the V-band the periods were more in
accordance to RRC group and the shape, amplitude characteristics
are in accordance with that.
\vspace{.1cm}
\par
\begin{figure*}
 \includegraphics[width=18cm]{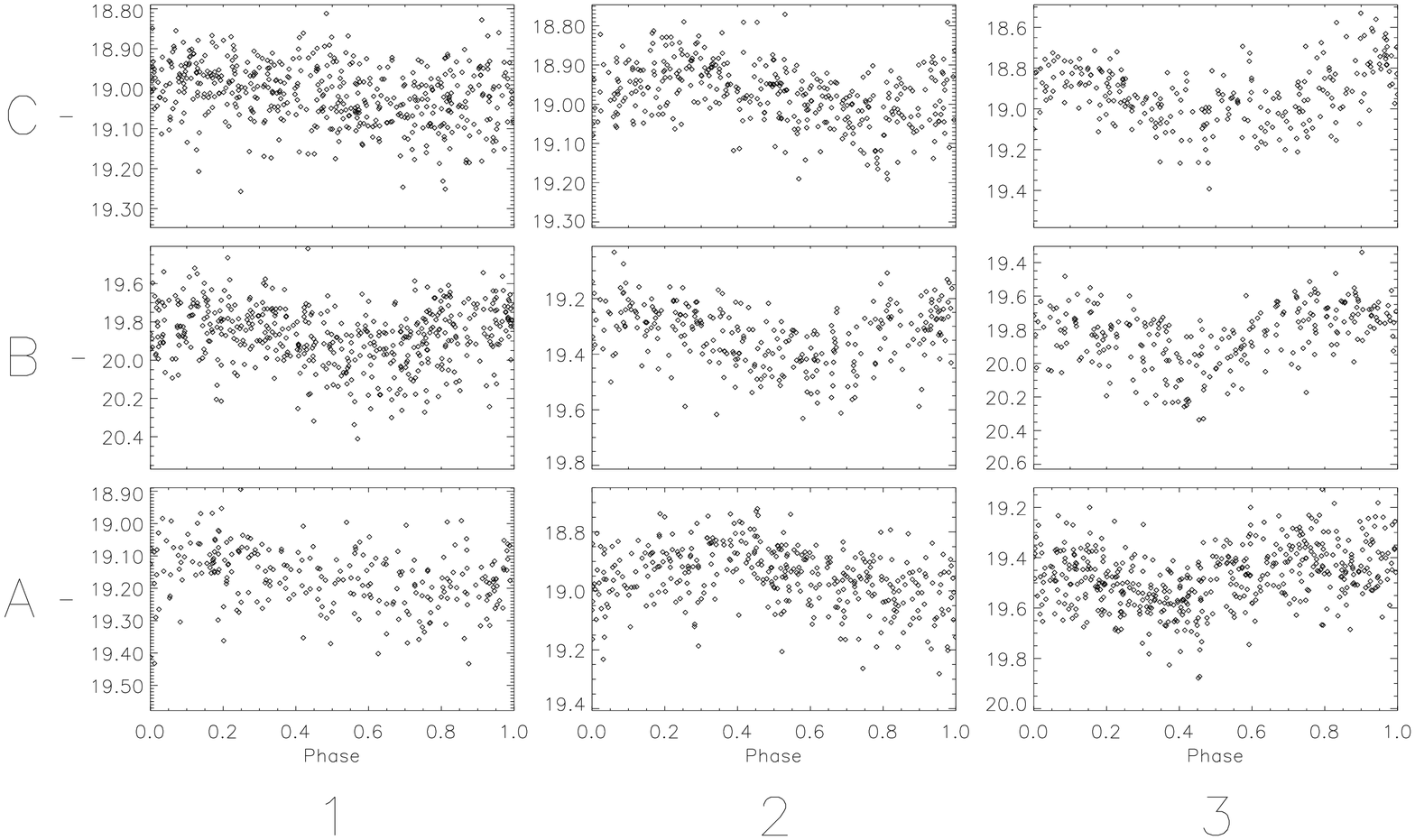}
\caption{Light-curves with the lowest  measure of similarity from
the {\bf OGLE RRL} catalog. Only the {\bf OGLE} I
band is shown.} \label{fig:OGLE_rrl_outliers}
\end{figure*}

 \noindent In the OGLE \rr~ catalog we identified three light-curves that
likely do not belong to this catalog.  Light-curve A1 of Fig.~\ref{fig:OGLE_rrl_outliers}
does not look periodic and the quoted period and
amplitude do not correspond to a typical \rr. Light-curve C1 has
quoted period of 0.86 days and amplitude of $<0.1$  in the I-band and
hard to make out signal. C3 is a light-curve that has period of
0.55 days thus most likely belonging to RRAB group but the
light-curve is very symmetric thus belonging to the RRC group.
This is one of the light-curves on which further investigation should
be performed.

\vspace{.3cm} \par \noindent \underline{EBs:}
Eclipsing Binary  stars are not due to physical variation but rather
due to occultation: one member of the pair of stars passes in front
of the other.
\vspace{.1cm}
\par
\begin{figure*}
 \includegraphics[width=17cm]{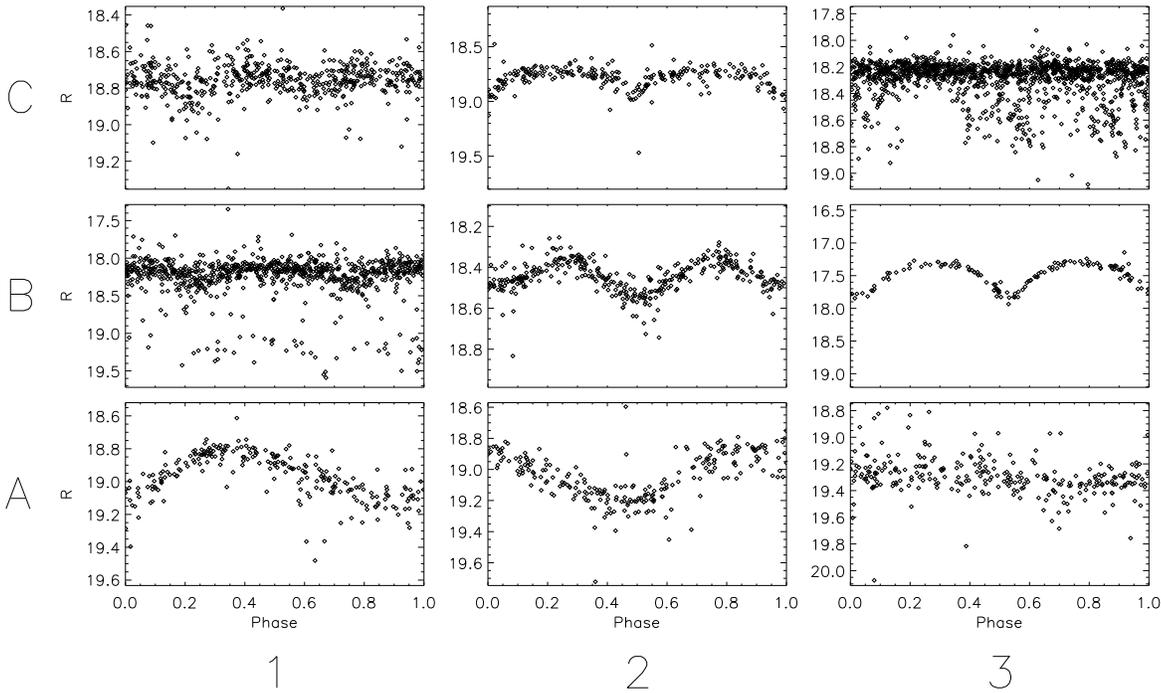}
\caption{Light-curves with the lowest  measure of similarity from
the {\bf MACHO EB} catalog. Only the {\bf MACHO} R
band is shown.} \label{fig:MACHO_EB_outliers}
\end{figure*}

 \noindent MACHO EB catalogs are submitted for publication in \cite{faccioli05}.
We used the method presented in this paper to help free the
submitted catalogs from outliers. We found few cases of outliers
that are shown here but will not be in the final published
catalogs. These are the light-curves shown in  Fig.~\ref{fig:MACHO_EB_outliers}-A1, A2, and A3
where all  three light-curves have  a symmetric
single occultation and periods consistent more with \rrs~ rather
than EBs. Light-curve in B1 shows no periodicity however after
examining the  V-band we were convinced that it is a true EBs.
The Light-curve shown in C3 shows a very noisy light-curve but after
cross correlating with the OGLE catalog we established that is a
proper EBs.

\vspace{.1cm}
\par \noindent In the OGLE EBs catalog most outliers are EBs
with very eccentric orbits thus appear as outliers since the
second minimum will  rarely be aligned with the second minimum of the
rest of the light-curves. However light-curve shown in panel C2 is
not a typical EB. There is either a 3rd body present in the
system producing a second occultation or some form of atmospheric
variation in one of the stars is synchronized with the binary system.
Perhaps there is a large reflection effect. This occurs when the
side of the dimmer star that is facing the Earth is illuminated by the brighter
companion star thus increasing the luminosity of the system  \citep{Pollarco93}.  This effect also includes radiative brightening.
For example the system could be a small hot star
with a much cooler sub-giant or giant component.
This light-curve warrants further investigation.

\par The reason why the algorithm identifies highly eccentric EBs as
outliers is well understood. At the same time it is well
understood that this is an indication that cross-correlation may
not be the best choice of similarity measure. In cases like these
a different measure of similarity must be employed.  These and
other potential  extensions will be investigated in future works.

%

% **************************************************
% **************************************************
%       RESULTS MACHO CEPH
% **************************************************
% **************************************************

\begin{table*}
\caption{ {\bf MACHO Cepheids} outliers} \label{tab:outliers_macho_ceph}
\begin{tabular}{lccccccl}
  \hline \hline
Survey &  Type  & ID & Plot COORD & Period [days] & Days of Obs & Num Obs &  Interpretation  \\
\hline \hline
MACHO & Ceph & 14.9223.221 & A1 &  242.49801 & 2721.04 &  883 & Multiple period. \\
MACHO & Ceph & 9.4511.14  & A2 &  12.19121 & 2720.82 & 595& Multiple period. \\
MACHO & Ceph & 4.7459.14  & A3 &   6.85445  & 2718.04 & 278 & \bf{EB}  \\
MACHO & Ceph & 60.7467.9  & B1 &   2.00174   & 2717.75 & 273& \bf{Periodic but unlikely }  \\
 \multicolumn{7}{}{}  &\bf{to be Cepheid} \\
MACHO & Ceph & 81.9490.26  & B2 &   1.14535   & 2715.86 &  204 & \bf{Blue band suggests an EB}  \\
MACHO & Ceph & 61.8562.27  & B3 &   7.33385    & 2715.84 & 366 & Multiple period.\\
MACHO & Ceph & 20.4309.2977 & C1 &    0.70794    & 2715.71 & 241& \bf{Not periodic/variable.} \\
MACHO & Ceph & 77.7067.41  & C2 &   8.00520    & 2709.83&  1333 & Multiple period. \\
MACHO & Ceph & 79.4659.3452  & C3 &    6.96615    & 2708.91  & 1352 & Multiple period. \\
\hline
\end{tabular}
\end{table*}

%\clearpage
%\begin{figure*}
% \includegraphics[width=18cm]{outliers_macho_ceph_bctrp.eps}
%\caption{Light-curves with the lowest  measure of similarity from
%the {\bf MACHO Cepheid} catalog. Shown here is only the BLUE
%band.} \label{fig:machoceph}
%\end{figure*}
\clearpage
% **************************************************
% **************************************************
%       RESULTS OGLE CEPH
% **************************************************
% **************************************************

\begin{table*}
\caption{ {\bf OGLE Cepheids} outliers} \label{tab:outliers_ogle_ceph}
\begin{tabular}{lcccccccl}
  \hline \hline
Survey &  Type  & Field & Number  & Plot COORD & Period [days] & Days of Obs & Num Obs &   Interpretation   \\
\hline \hline
OGLE-II & Ceph & 17 &70123   & A1 &  28.96683 & 1419 &  105 & \bf{Not enough/noisy data.} \\
OGLE-II & Ceph & 13 &184117  & A2 &  13.64083 & 1419 & 245 & \bf{Atypical asymmetry.}  \\
OGLE-II & Ceph & 21 &40876   & A3 &  4.97338 &  1418 & 248 & \bf{Needs further study.}\\
OGLE-II & Ceph & 17 &221134  & B1 &  11.22865 & 1418 & 243&  \\
OGLE-II & Ceph & 21 &119037  & B2 &  0.87813 & 1417 &  264 &   \\
OGLE-II & Ceph & 14 &114046  & B3 &  0.9094 &  1415 & 238 & \\
OGLE-II & Ceph & 18 &185847  & C1 &  12.20018 & 1414 & 244&  \\
OGLE-II & Ceph & 4  &168269   & C2 &  0.72923 &1417 &  327 & \\
OGLE-II & Ceph & 4  &427313   & C3 &  0.67413 & 1414 & 454 &\\
\hline
\end{tabular}
\end{table*}
\clearpage

% **************************************************
% **************************************************
%       RESULTS MACHO RRL
% **************************************************
% **************************************************

\begin{table*}
\caption{ {\bf MACHO RRL} outliers} \label{tab:outliers_macho_rrl}
\begin{tabular}{lccccccl}
  \hline \hline
Survey &  Type  & ID & Plot COORD & Period [days] & Days of Obs & Num Obs &   Interpretation  \\
\hline \hline
MACHO & RRL &  48.2992.463  & A1 &  0.11828 & 2720.99 &  151  & RRC.Incorrect period in R-band. \\
\multicolumn{7}{}{} &  $P_V=0.35484$.  \\
MACHO & RRL &  82.8772.705  & A2 & 0.98011  & 2717.79 &  747 & \bf{Unlike RRL.}   \\
MACHO & RRL &  73.13488.41 & A3 &  0.53464 &2716.77 &  121 &   \bf{Unlike RRL.}  \\
MACHO & RRL &  76.10942.176 & B1 &  0.46759 & 2714.75 &  121 &   RRC. Incorrect period in R-band. \\
 \multicolumn{7}{}{} &   $P_V= 0.35042$.  \\
MACHO & RRL &  79.5499.2627 & B2 &  0.25465 & 2710.73 &  1387 & RRC. Incorrect period in R-band. \\
 \multicolumn{7}{}{} &  $P_V= 0.33756$.\\
MACHO & RRL &  67.10489.79  & B3 &  0.29736 & 2707.9 &  273 &  \\
MACHO & RRL &  34.9080.261 & C1 &  0.30795 & 2700.79 &   156&  RRC. Incorrect period in R-band. \\
 \multicolumn{7}{}{} &  $P_V= 0.46193$.  \\
MACHO & RRL &  37.6316.471  & C2 &  0.62069 & 2697.96 &  125 &  \\
MACHO & RRL &  49.6623.336   & C3 &  0.62001 & 2689.9 & 178  & \\
\hline
\end{tabular}
\end{table*}
\clearpage

%\begin{figure*}
% \includegraphics[width=18cm]{outliers_macho_rrl_blue_ctrp.eps}
%\caption{Light-curves with the lowest  measure of similarity from
%the {\bf MACHO RRL} catalog. Shown here is only the V band.}
%\label{fig:MACHO_rrl_outliers}
%\end{figure*}
\clearpage
% **************************************************
% **************************************************
%       RESULTS OGLE RRL
% **************************************************
% **************************************************

\begin{table*}
\caption{{\bf OGLE RRL} outliers} \label{tab:outliers_ogle_rrl}
\begin{tabular}{lccccccl}
  \hline \hline
Survey &  Type  & ID(RA-DEC) & Plot COORD & Period [days] & Days of Obs & Num Obs &  Interpretation  \\
\hline \hline
OGLE & RRL &  053803.42-695656.4  & A1 &  0.3323824 & 1420 &  267  &  \bf{Unknown; Noisy data.}\\
OGLE & RRL &  053325.94-701109.8  & A2 &  0.2876012  & 1420 &  371 &   \\
OGLE & RRL &  052447.86-694319.0  & A3 &  0.2585634  & 1420 &  495 &\\
OGLE & RRL &  052436.03-694541.8  & B1 &  0.2232339  & 1420 &  504 &   \\
OGLE & RRL &  053525.67-702210.2  & B2 &  0.2164212  & 1420 &  298 &   \\
OGLE & RRL &  054036.89-701424.8  & B3 &  0.2361195 & 1420 &  268 &  \\
OGLE & RRL &  052219.98-691907.1  & C1 &  0.8616601 & 1420 &  503 &   \bf{Unknown.} \\
 \multicolumn{7}{}{} &  \bf{RRAB period but amplitude too small.}    \\
OGLE & RRL &  053241.91-702718.9  & C2 &  0.2749622 & 1420 &  373 &  \\
OGLE & RRL &  054609.21-702316.7  & C3 &  0.5494448 & 1420 &  263  &  \bf{Unknown.} \\
 \multicolumn{7}{}{} &  \bf{RRAB period but symmetric.} \\
\hline
\end{tabular}
\end{table*}
\clearpage

% **************************************************
% **************************************************
%       RESULTS MACHO EB
% **************************************************
% **************************************************
\begin{figure*}
\begin{minipage}{7cm}
   \includegraphics[width=6cm]{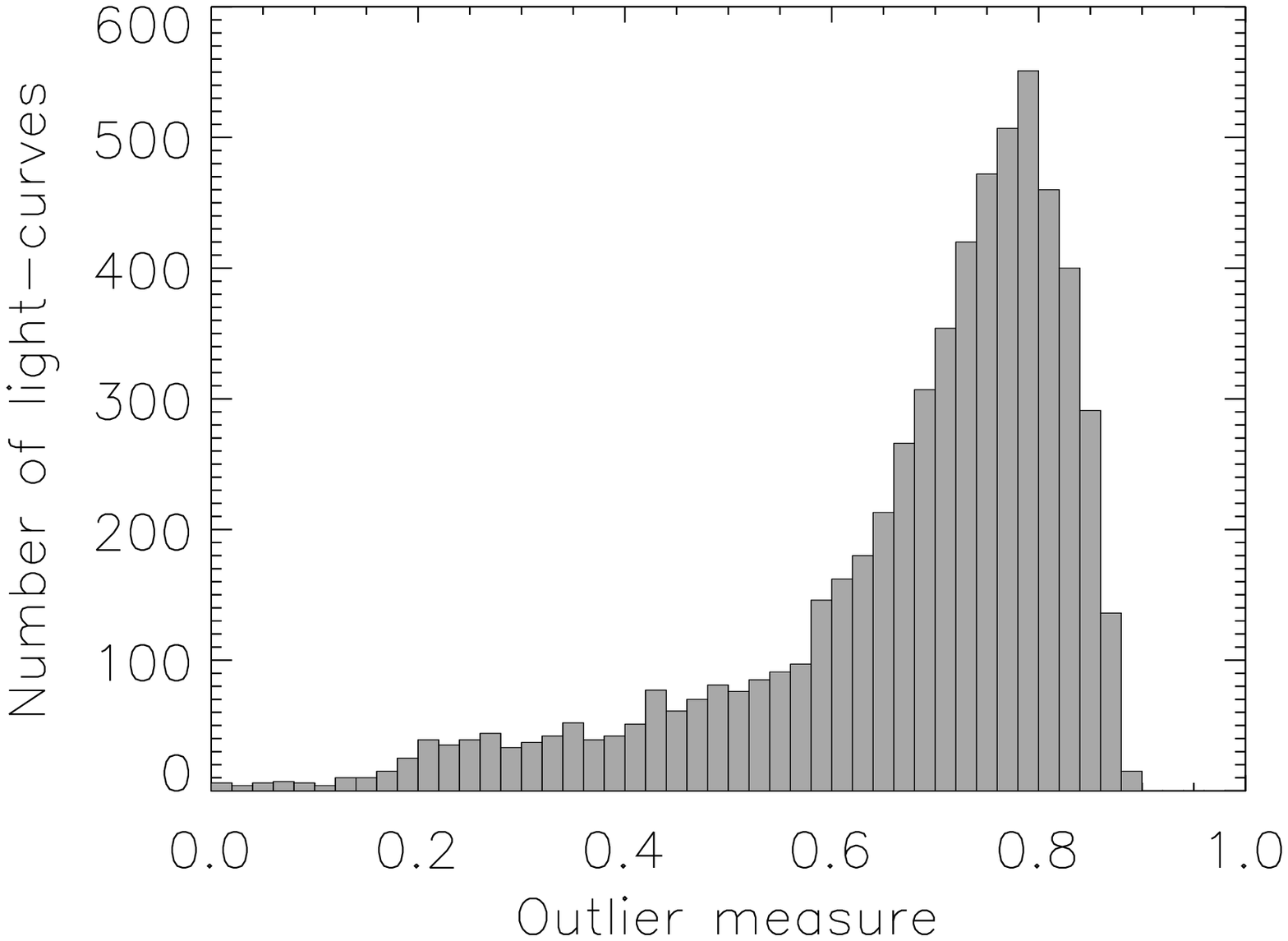}
  \caption{ {\small Histogram of the outlier measure for 6064 {\bf EBs}  in the
{\bf MACHO} sample.}}
  \label{fig:MACHO_EB_hist}
\end{minipage}
\hspace{1cm}
\begin{minipage}{7cm}
    \includegraphics[width=6cm]{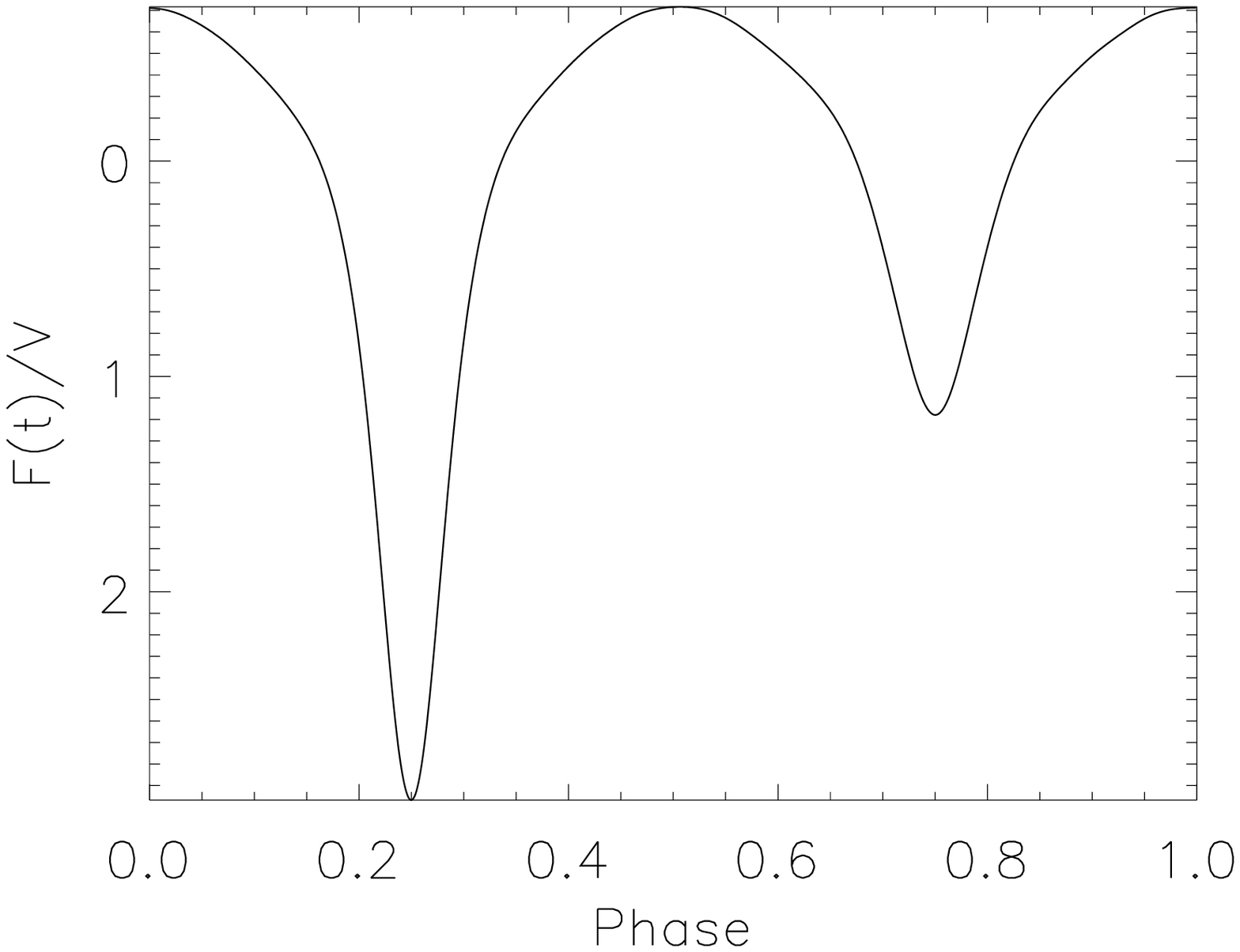}
    \caption{ {\small Centroid light-curve  for 6064 {\bf EBs}  in the
{\bf MACHO} sample.}}
\end{minipage}
\end{figure*}

\begin{table*}
\caption{ {\bf MACHO EB} outliers} \label{tab:outliers_macho_eb}
\begin{tabular}{lccccccl}
  \hline \hline
Survey &  Type  & ID & Plot COORD & Period [days] & Days of Obs & Num Obs &  Interpretation  \\
\hline \hline
MACHO & EB & 64.7964.375   & A1 &  0.32279 & 2728.73 &  263 & {\bf RRAB.} \\
 \multicolumn{7}{}{} &  \bf{Asymmetry, period. }\\
MACHO & EB & 68.10485.363   & A2 &  0.36804  & 2714.83 &  205 & {\bf RRAB.} \\
 \multicolumn{7}{}{} &  \bf{Asymmetry, period.}\\
MACHO & EB & 27.10782.248   & A3 &  0.28717 & 2713.96 &  294 & {\bf RRAB} \\
 \multicolumn{7}{}{} &  \bf{Asymmetry, period.}\\
MACHO & EB & 212.15797.121   & B1 &  0.67719 & 2711.93 &  910 &  Red band is noisy. Blue band is OK.\\
MACHO & EB & 25.3836.269  & B2 &  2.27436 & 2711.82 &  341 & \\
MACHO & EB & 36.7395.92  & B3 &  0.31633  & 2702.76 &  276 &\\
MACHO & EB & 22.4871.431  & C1 &  3.04861 & 2702.73 &  530 & \\
MACHO & EB & 57.4953.114  & C2 &  1.25421 &  2660.68 & 278 & \\
MACHO & EB & 80.7194.423   & C3 &  2.60078 & 2649.06 &  1370 &  \\
\hline
\end{tabular}
\end{table*}
\clearpage

%\begin{figure*}
% \includegraphics[width=18cm]{outliers_macho_eb_bctrp.eps}
%\caption{Light-curves with the lowest  measure of similarity from
%the {\bf MACHO EB} catalog. Shown here is only the {\bf MACHO} V
%band.} \label{fig:machoceph}
%\end{figure*}
\clearpage

% **************************************************
% **************************************************
%       RESULTS OGLE EB
% **************************************************
% **************************************************
\begin{figure*}
\begin{minipage}{7cm}
   \includegraphics[width=6cm]{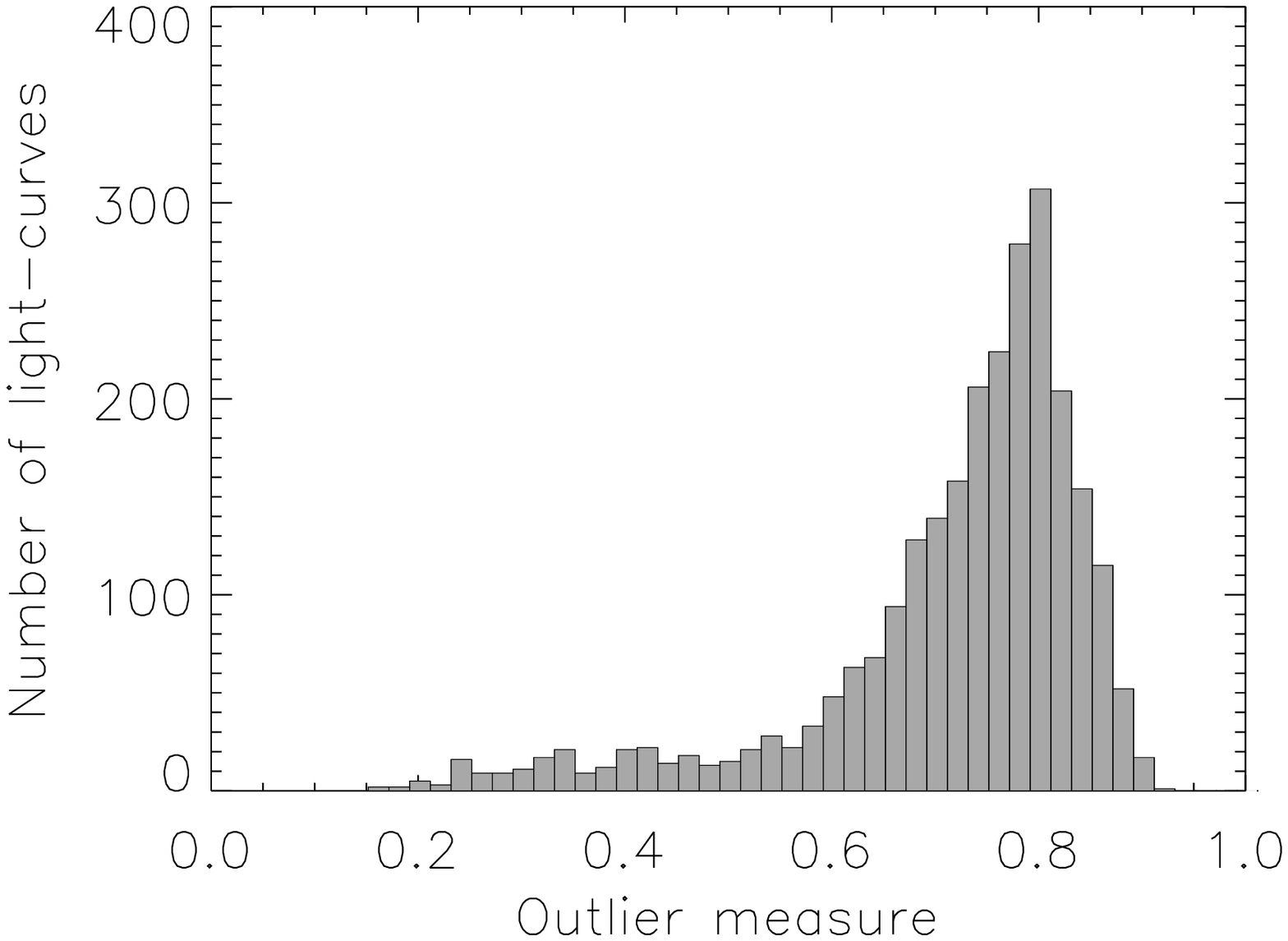}
  \caption{ {\small Histogram of the outlier measure for 2580 {\bf EBs}  in the
{\bf OGLE} sample.}}
  \label{fig:OGLE_EB_hist}
\end{minipage}
\hspace{1cm}
\begin{minipage}{7cm}
    \includegraphics[width=6cm]{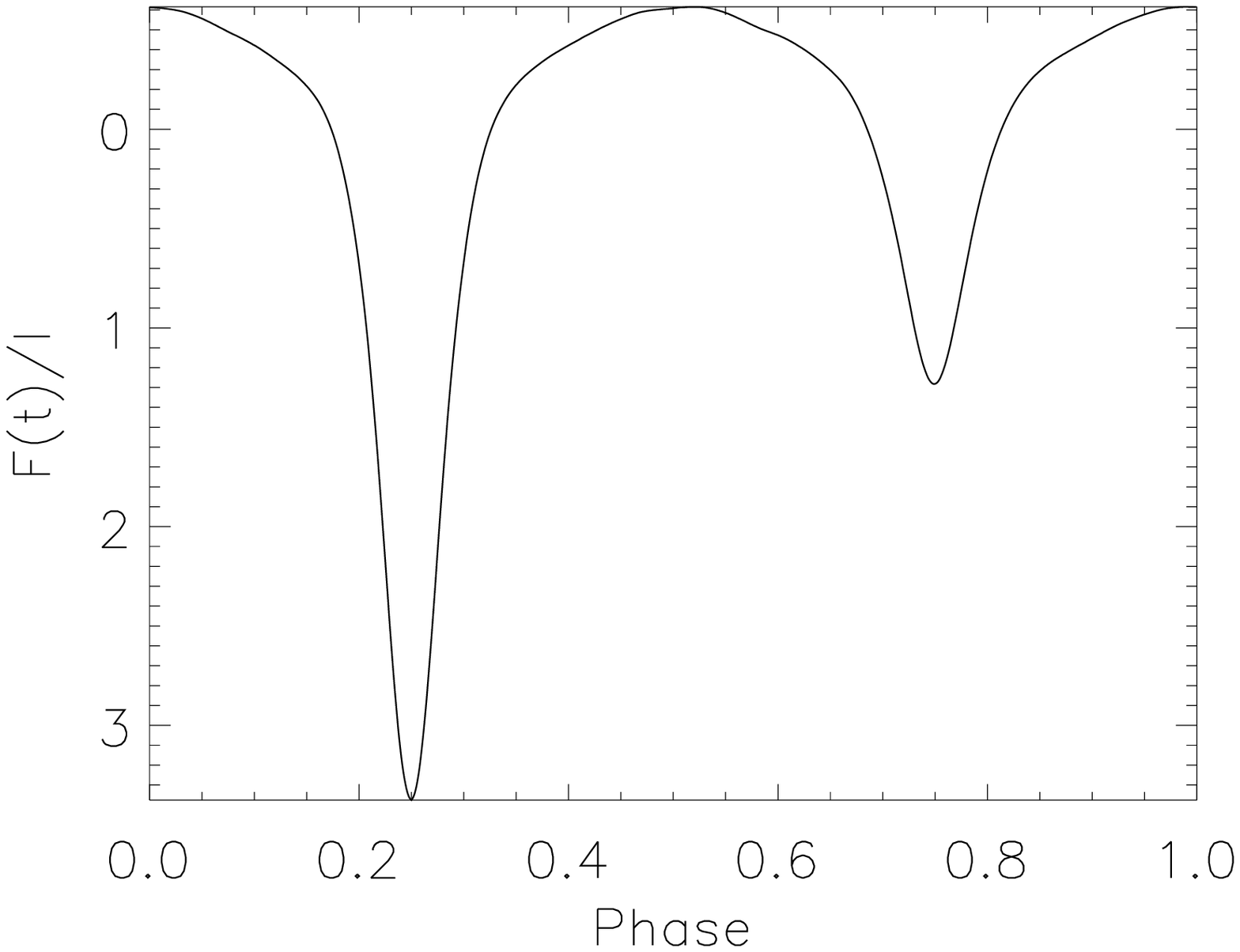}
    \caption{ {\small Centroid light-curve  for 2580 {\bf EBs}  in the
{\bf OGLE} sample.}}
\end{minipage}
\end{figure*}

\begin{figure*}
 \includegraphics[width=18cm]{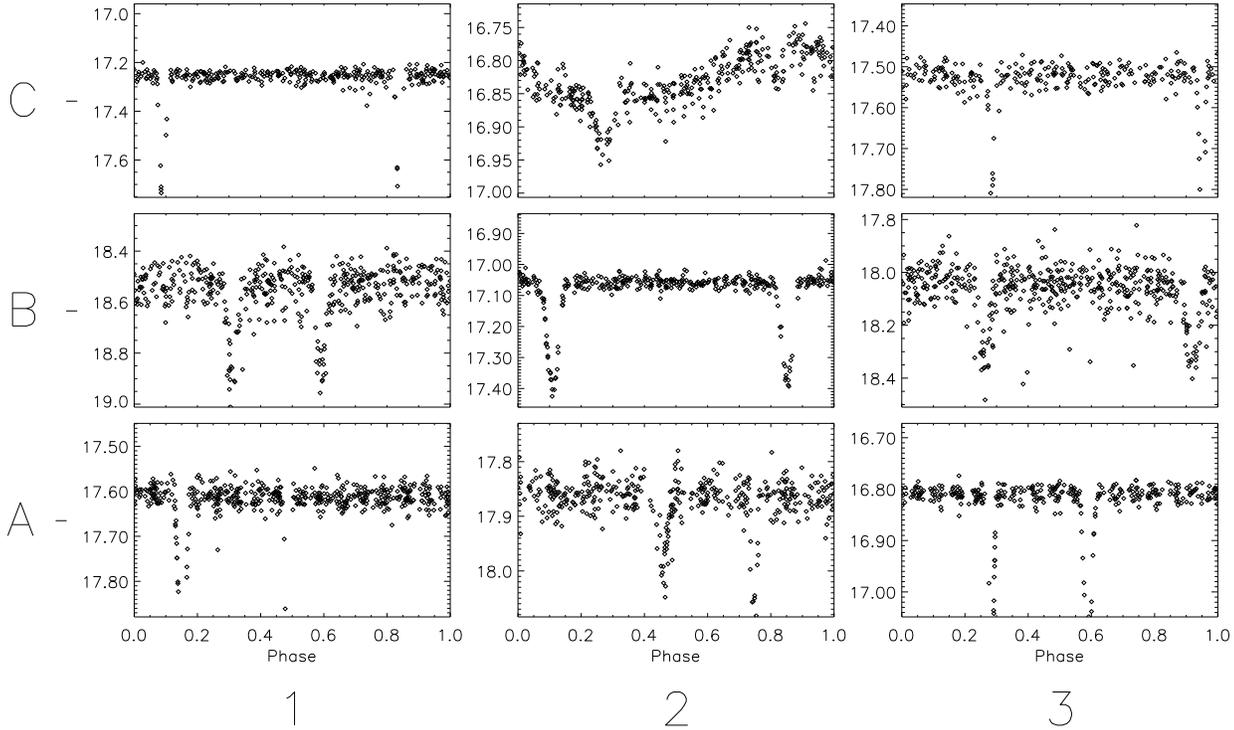}
\caption{Light-curves with the lowest  measure of similarity from
the {\bf OGLE-II EB} catalog. Shown here is only the {\bf OGLE} I
band is shown.}  \label{fig:OGLE_EB_outliers}
\end{figure*}

\begin{table*}
\caption{ {\bf OGLE EB} outliers} \label{tab:outliers_ogle_eb}
\begin{tabular}{lccccccl}
  \hline \hline
Survey &  Type  & ID(RA-DEC) & Plot COORD & Period [days] & Days of Obs & Num Obs &  Interpretation  \\
\hline \hline
OGLE-II & EB & 052937.78-700903.4   & A1 &  15.03314 & 1239 &  503 & Eccentric orbit.\\
OGLE-II & EB & 051915.79-693808.1   & A2 &   8.03376 & 1238 &  432 & Eccentric orbit.\\
OGLE-II & EB & 051519.31-692640.3   & A3 &   15.96256 & 1235 &  360 & Eccentric orbit.\\
OGLE-II & EB & 051858.34-693946.4   & B1 &  2.29555 & 1235 &  473 &  Eccentric orbit.\\
OGLE-II & EB & 051700.39-691813.8  & B2 &  5.29129 & 1235 &  368 & Eccentric orbit.\\
OGLE-II & EB & 052521.32-694858.9  & B3 &  4.12088 & 1234 &  500 & Eccentric orbit.\\
OGLE-II & EB & 051734.54-692736.5  & C1 &  14.58252 & 1234 &  325 & Eccentric orbit.\\
OGLE-II & EB & 051657.87-690328.1  & C2 &  5.66141 & 1238 &  365 & \bf{EB with reflection effect.} \\
OGLE-II & EB & 050646.85-683700.4   & C3 &  12.14988 & 1420 &  264 &  Eccentric orbit. \\
\hline
\end{tabular}
\end{table*}
\clearpage

\section{Future Work}
This paper is not intended to study all possible methods for
finding outliers in datasets of light-curves  but rather to help
demonstrate and hopefully convince others how an automatic method like
this  can be applied to facilitate the discovery of new,
interesting variable objects. Special emphasis should be given to
the choice of measure of similarity. An attempt to study this
issue will be made in a second paper where we will study how to
employ more than one measure of similarity.

In this paper we have used particular preprocessing tools and we
tweaked our preprocessing steps for each catalog.  We are planning
a full released of the software which will include many
preprocessing options  and optimized algorithms as a downloadable
software and as an on-line tool and web services in the near future
({\tt http://darwin.cfa.harvard.edu/LightCurves/s/}).

%============================================================================
%=========================================================================
%      CONCLUSIONS
%============================================================================
%============================================================================

\section{Conclusions }

In this paper we presented a methodology based on
cross-correlation as a measure of similarity that enables us to
discover outliers in catalogs of periodic light-curves. We
established the methodology in Fourier space and extended the
cross-correlation to accommodate observational errors.

The  results from the application of our method on catalogs of
classified periodic stars from the MACHO and OGLE projects are
encouraging, and establish that our method correctly identifies
light-curves that do not belong to these catalogs as outliers.

We have identified light-curves that were simply misclassified,
light-curves that were folded with the wrong period and so appear
different, and light-curves that emerged as unique.

We show how with careful approximations our method can be applied
to very large catalogs thus making it a useful tool for the
upcoming new surveys Pan-STARRS  ({\tt http://pan-starrs.ifa.hawaii.edu})
and LSST ({\tt http://www.lsst.org}).

We have nonetheless also concluded that a single measure of
similarity is not adequate to capture all features for all types
of light-curves and we understand that an extension of our method
that utilizes more measures (comparison of Fourier components,
wavelet coefficients etc) or combinations of measures  has to be
carried out; these  will be presented in a future paper.

It is worth mentioning that other works performing automated
classification  of  light-curves (\cite{Brett2004MNRAS}) can also,
in principle, find outliers. However since their focus is
classification there is no guarantee that an outlier will be
identified. This is because a light-curve must be clearly
decoupled from all  clusters in order to be considered as an
outlier where in our case, since we do not have clusters, any
light-curve can be classified as an outlier. This distinction is
important in order to appreciate the advantage of our method.
Moreover a classification method cannot scale as $N$ whereas our
method can do so in some approximation schemes.

We would like to make one last point.  The situation of datasets that
are not fully processed is going to become more common as the larger surveys come on-line.
In the near future it will become nearly impossible to fully ``clean''
datasets without the use of automated methods such as the one
presented here.  We believe we have shown that our method has
great utility at a number of steps along the processing pipeline.

%\section{References}

\section{acknowledgments}
This work uses public domain data from the MACHO project whose
work was performed under the joint auspices of the U.S. Department
of Energy, National Nuclear Security Administration by the
University of California, Lawrence Livermore National Laboratory
under contract No. W-7405-Eng-48, the National Science Foundation
through the Center for Particle Astrophysics of the University of
California under cooperative agreement AST-8809616, and the Mount
Stromlo and Siding Spring Observatory, part of the Australian
National University. This work also uses public domain data
obtained by the OGLE project.
We thank Kem Cook, Doug Welch and Gabe Prochter for 
compiling the lists of potential \rr~ and Cepheids 
and to Kem Cook for providing MACHO to standard magnitude
transformations.  We also thank Edward Guinan for his insight into some of
the outlier light-curves.
\bibliography{LightCurveMining}        % Light-CurveMining.bib is the name of our database

% ==================================================
% ==================================================
% APPENDIX
% ==================================================
% ==================================================

\begin{appendix}

% ----------------------------------------------------
% APPENDIX - ERROR PROPAGATION
% ----------------------------------------------------

\section{Convolution in Fourier Space} \label{app:conv}
Let $x(n)$ and $y(n)$ be arbitrary functions of discrete time n with Fourier transforms.
Take
\begin{eqnarray}
  x(n) = \ift \left[ {\cal X} (\nu) \right](n) &=& \frac{1}{N} \sum_{n=0}^{N-1} {\cal X} (\nu) e^{2\pi i\nu n /N} \\
    & =& \frac{1}{N} \sum_{n=0}^{N-1} \bar{{\cal X}} (\nu) e^{-2\pi i\nu n /N}  ,  \\
  y(n) = \ift \left[ {\cal Y} (\nu) \right](n) &=& \frac{1}{N} \sum_{n=0}^{N-1} {\cal Y} (\nu) e^{2\pi i\nu n /N}
   \\
    & =& \frac{1}{N} \sum_{n=0}^{N-1} \bar{{\cal Y}} (\nu) e^{-2\pi i\nu n /N}  ,
\end{eqnarray}
where $\bar{{\cal X}}$ and $\bar{{\cal Y}}$ are the complex conjugates Fourier transforms and
 $\ift(n)$ is the inverse Fourier transform. The correlation given a time lag $\tau$

\begin{equation}
  r^2_{xy}(\tau)  =   \sum_{n=0}^{N-1} x(n) \, y(n-\tau) ,
\end{equation}
is
\begin{eqnarray}
       r^2_{xy}(\tau)         &=&  \sum_{n=0}^{N-1} x(n) \frac{1}{N}
\sum_{\nu=0}^{N-1} \bar{{\cal Y}}(\nu)
                                          e^{i 2 \pi (\tau-n) \nu /N}
\nonumber \\
            &=&   \frac{1}{N} \sum_{\nu=0}^{N-1} \bar{{\cal Y}}(\nu) e^{
i 2 \pi  \nu \tau /N }\sum_{n=0}^{N-1}
                      x(n)  e^{- i 2 \pi \nu n /N} \nonumber \\
        &=&   \frac{1}{N} \sum_{\nu=0}^{N-1}  \bar{{\cal Y}}(\nu)  {\cal
X}(\nu)  e^{ i 2 \pi  \tau \nu /N }      \nonumber    \\
          &=&    {\cal F}^{-1}  \left[    \bar{{\cal Y}}(\nu)  {\cal
X}(\nu)    \right] (\tau)\: .
\end{eqnarray}

% ----------------------------------------------------
% APPENDIX - ERROR PROPAGATION
% ----------------------------------------------------
\section{Error propagation}
\label{app:errors}
The SG smoothing can be written as a simple linear sum over neighboring
points

\begin{equation}
y_{s}=\sum_{i=-\frac{N-1}{2}}^{\frac{N-1}{2}}C_{i} \: y_{i} \: ,
\end{equation}
where the coefficients $C_i$ are the difficult thing to deduce, but have
no errors in them (they do not depend on the data).
The error in the smoothed value is then given by,

\begin{equation}
\sigma_{y_{s}}=\sqrt{ \sum \left( \frac{ \partial y_s }{\partial y_i} \:
\sigma_{y_{i}} \right)^2 } \:,
\end{equation}

\noindent implying,

\begin{equation}
\sigma_{y_{s}}=\sqrt{ \sum \left( C_i \: \sigma_{y_{i}} \right)^2 } \: .
\end{equation}

\vspace{1cm}
\noindent To get the value of a measurement $y$ for a given $x$
using linear interpolation between the two points $(x_1,y_1)$ and
$(x_2, y_2)$ we have,

\begin{equation}
\label{Equ:Line}
y \,\, = \,\,   \eta \, y_2  \,+\,(\eta -1 ) \, y_1 \: ,
\end{equation}
where $\eta$ is defined as:

\begin{equation}
\label{Equ:fit1}
\eta = \frac{x-x_1}{x_2-x_1} \;.
\end{equation}

\noindent Using the rules of error propagation,

\begin{equation}
\sigma_{y}=
\sqrt{  \left( \frac{ \partial y }{\partial y_1} \, \sigma_{y_1} \right)^2
\,+\,
\left( \frac{ \partial y }{\partial y_2} \, \sigma_{y_2} \right)^2
} \: ,
\end{equation}
calculating the derivatives we find,

\begin{equation}
\sigma_{y}=
\sqrt{  (1-\eta)^2 \, \sigma_{y_1}^2
\,+\,
 \eta^2 \, \sigma_{y_2}^2
} \: .
\end{equation}

\vspace{1cm}
\noindent Similarly we can estimate the errors for the running averages where
the running averages are:
\begin{equation}
  y = \sum_{i \in \mathrm{window}} e^{ -\frac{(y-y_i)^2}{2 \omega^2}} y_{i}  \; ,
\end{equation}
where $\omega$ is the window size.
Estimating the derivatives we get
\begin{equation}
  \sigma_{y}^2 = \sum_{i \in \mathrm{window} } e^{-\frac{ (y-y_{i})^2 }{   \, \omega^2  } } \;
  \left[ 1-  \frac{ \, y_i \, (y-y_i) }{ \omega} \right]^2  \sigma_{y_i}^2
\end{equation}

\end{appendix}

\end{document}